\documentclass[nohyper,notoc]{article} 
\usepackage{graphicx}
\input epsf

\def\beq{\begin{equation}}
\def\eeq{\end{equation}}
\def\beqa{\begin{eqnarray}}
\def\eeqa{\end{eqnarray}}
\def\bea{\begin{eqnarray}}
\def\eea{\end{eqnarray}}
\def\bq{\begin{quote}}
\def\eq{\end{quote}}
\def \lsim{\mathrel{\vcenter
     {\hbox{$<$}\nointerlineskip\hbox{$\sim$}}}}
\def \gsim{\mathrel{\vcenter
     {\hbox{$>$}\nointerlineskip\hbox{$\sim$}}}}
\def\gappeq{\mathrel{\rlap {\raise.5ex\hbox{$>$}}
{\lower.5ex\hbox{$\sim$}}}}
\def\lappeq{\mathrel{\rlap{\raise.5ex\hbox{$<$}}
{\lower.5ex\hbox{$\sim$}}}}

\def\m3e{\mu \to e \bar{e} e}
\def\a{\alpha}
\def\b{\beta}

\def\lam{\lambda}

\def\ifmath#1{\relax\ifmmode #1\else $#1$\fi}
\def\lsim{\mathrel{\raise.3ex\hbox{$<$\kern-.75em\lower1ex\hbox{$\sim$}}}}
\def\gsim{\mathrel{\raise.3ex\hbox{$>$\kern-.75em\lower1ex\hbox{$\sim$}}}}

\def\eq#1{eq.~(\ref{#1})}

\def\anti{\overline}

\def\cbma{c_{\beta-\alpha}}

\def\sbma{s_{\beta-\alpha}}

\def\phm{\phantom{-}}

\def\ifmath#1{\relax\ifmmode #1\else $#1$\fi}

\def\call{\mathcal{L}}
\def\s2W{\sin^2\theta_W}

\def\sbmaii{s^2_{\beta-\alpha}}
\def\cbmaii{c^2_{\beta-\alpha}}

\def\hl{h}
\def\ha{A}
\def\hh{H}

\def\mw{m_W}

\def\ls#1{\ifmath{_{\lower1.5pt\hbox{$\scriptstyle #1$}}}}
\def\lss#1{\ifmath{^{\,\lower2.5pt\hbox{$\scriptstyle #1$}}}}

\def\half{\ifmath{{\textstyle{1 \over 2}}}}

\def\m{\mu}

\def\tmg{\tau \to \mu \gamma}

\begin{document}
\renewcommand{\thefootnote}{\fnsymbol{footnote}}
\begin{center}
{\Large {\bf 
 Lepton flavour violating Higgs and $\tau \to \mu \gamma$ }
}
\vskip 25pt
{\bf 
{ Sacha Davidson} \footnote{E-mail address:
s.davidson@ipnl.in2p3.fr}
and {  Gerald  Grenier} \footnote{E-mail address:
g.grenier@ipnl.in2p3.fr}
}
\vskip 10pt
{\it  IPNL, Universit\'e de Lyon, Universit\'e Lyon 1, CNRS/IN2P3, 4 rue E. Fermi 69622 Villeurbanne cedex, France
}

\vskip 20pt
{\bf Abstract}
\end{center}

\begin{quotation}
  {\noindent\small
We update phenomenological constraints on  a Two Higgs Doublet Model
with lepton flavour non-conserving Yukawa couplings. We review that
$\tan \beta$ is ambiguous in such ``Type III''
models, and define it from the $\tau$ Yukawa coupling.
The neutral
scalars $\phi$
 could be searched for at hadron colliders in $ \phi \to \tau \bar{\mu}$,
and are constrained by 
the rare decay $\tau \to \mu \gamma$.
The Feynman  diagrams for the
collider process, with Higgs production via
gluon fusion, are similar to    the two-loop ``Barr-Zee'' diagrams
which contribute to $\tau \to \mu \gamma$. 
Some ``tuning'' is required to
obtain a  collider cross-section of order the Standard Model
expectation for $\sigma (gg \to h_{SM} \to \tau^+ \tau^-)$,
while agreeing  with the current bound from $\tau \to \mu \gamma$. 
 

\vskip 10pt
\noindent
}

\end{quotation}

\vskip 20pt  

\setcounter{footnote}{0}
\renewcommand{\thefootnote}{\arabic{footnote}}

\section{Introduction}

The Two Higgs Doublet Model (2HDM) may be the low energy
effective theory for many models of Beyond-the Standard Model (BSM) 
 physics  at the TeV scale. 
 Various  Higgses  could  be the first
signals of BSM physics  discovered at  hadron colliders.
The aim of this paper is therefore to study the implications
for collider searches, of precision physics bounds  
 on  a generic 2HDM  with lepton flavour
violating couplings.
Similar analyses have previously been performed in
\cite{Iltan:2001rp,Diaz:2000cm,Assamagan:2002kf,Arcelli:2004af,Kanemura:2005hr,Li:2008xx}.
 We assume that the 
additional Higgses are  the only BSM particles  
with masses $\lsim$ 400 GeV, and  consider  constraints
on the Higgs parameters which are comparatively independent
of additional New Physics at higher scales.

The 2HDM \cite{2hdm} (see \cite{HHG} for an introduction)
consists in adding a second Higgs doublet to the Standard
Model.  Despite being a fairly minimal  extension\footnote{The Higgs sector of
the Minimal Supersymmetric Standard Model  \cite{Djouadi:2005gj}
is a  2HDM. Larger Higgs sectors with multiple
doublets and singlets\cite{Barger:2009me,Grimus:2007if},  
and/or triplets, can also
be considered.}, it  has many variants. In particular,
a discrete symmetry \cite{disc} can be imposed on the Higgs plus fermion
Lagrangian, to avoid tree level flavour changing
neutral interactions. 
We focus here on    ``Type III'' models, meaning that
no discrete symmetries are present, so there is no
unique definition of $\tan \b$. We neglect non-renormalisable
operators
\cite{Delaunay:2007wb,Antoniadis:2009rn}
 in the potential. In the absence
of a discrete symmetry, the  fermions can
couple to both Higgs doublets with  generic
Yukawa matrices, which allows
flavour-changing  tree-level  couplings of
the physical Higgses. For simplicity, we make
the (unrealistic) assumption that our Higgses only  have 
{\it lepton} flavour violating interactions
\footnote{See, for instance, \cite{Atwood:1996vj} and
references thereto, for a discussion of quark
flavour changing interactions in type III models.}. 
We are particularily interested in the Higgs$-\tau -\mu$
interaction.

\begin{figure}[ht]
  \centerline{
 \scalebox{0.90}{\includegraphics{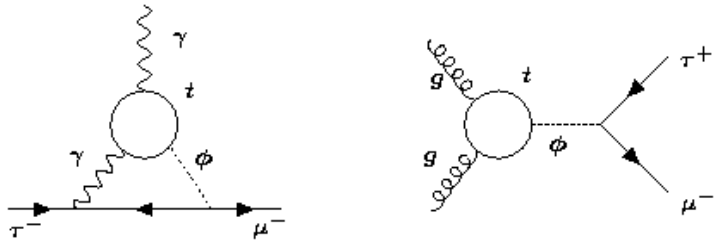}}}
   \caption{\small 
On the left, the ``Barr-Zee'' diagram which contributes to
 $\tau \to \mu \gamma$. On the right,  the production/decay diagram
for a neutral Higgs $\phi$ at  hadron collider.}
\label{bz}
\end{figure}

We emphasize  that, for a generic neutral
Higgs  $\phi$,  the search for $\phi \to \tau \bar{\mu}$
at hadron colliders\cite{Han:2000jz,Assamagan:2002kf,Arcelli:2004af}  
should take into account the
bounds  on $\tau \to \mu \gamma$\cite{Kanemura:2005hr}. In figure \ref{bz},
on the left is shown  a fermion
loop diagram of   ``Barr-Zee''  type
\cite{Bjorken:1977br,Barr:1990vd,Cheung:2003pw,El Kaffas:2006nt,ZW,Chang:1993kw} 
which contributes to
$\tau \to \mu \gamma$.  
Beside it is  the  similar diagram for
Higgs production and decay to $\tau \bar{\mu}$ at a
hadron collider. In the absence of cancellations
between the different neutral Higgses in the Barr-Zee,
it is clear that an upper bound on 
$BR(\tau \to \mu \gamma)$ sets a bound on
$\sigma (gg \to \phi \to \tau \bar{\mu})$.

There is a large literature on flavour
changing observables in the  Type
III 2HDM.  Various textures for the
Yukawa matrices are discussed in \cite{Carcamo:2006dp}
(see also references therein).
 Recently, the concept
of Minimal Flavour Violation has
been extended to multi-Higgs models \cite{Botella:2009pq}.
See also \cite{Pich:2009sp} for a clear discussion
of CP violation in  a Type III  2HDM that
has no tree-level flavour changing neutral couplings. 
In the Minimal Supersymmetric Standard Model
(MSSM), Higgs decay to $\tau^\pm \mu^\mp$ \cite{Brignole:2003iv},  
and its  relation to one loop rare $\tau$ decays
\cite{BR} have been extensively studied (see also
citations of \cite{Brignole:2003iv}).
 Our analysis differs from 
\cite{Iltan:2001rp,Assamagan:2002kf,Kanemura:2005hr}
in that we have included the Barr-Zee diagram
in our calculation of $\tmg$. This can be
relevant if the flavour-changing Higgs has  
${\cal O} (1)$  coupling to the top (``small
$\tan \beta''$). The current  bound
on $BR(\tmg)$ is also slightly stronger
that the value used   by \cite{Kanemura:2005hr},
who performed a  more
complete study of $\tau$ decay bounds
in 2005. Finally, we are attentive
to the definition of $\tan \beta$; 
we allow the scalar potential of the Higgs
to take  its most general renormalisable 
CP-conserving form,
which does not allow a Lagrangian-basis-independent
definition of $\tan \beta$. The unphysical character of $\tan \b$
in type III models has been treated in various ways by
previous authors, which makes it difficult to compare results.
We introduce a ``physical''  definition of $\tan \b$,
via  the Yukawa coupling of the $\tau$, to facilitate comparaison
with ``almost type II'' models such as the MSSM.

In section \ref{notn}, we introduce our 
``basis independent'' \cite{Davidson:2005cw,Ivanov:2005hg,Accomando:2006ga,Ivanov:2007de} notation for the  2HDM.
We  discuss electroweak 
precision bounds on the Higgs masses in section \ref{bds},
and rapidly review bounds  (for instance, on the Higgs potential) that
are secondary  in our analysis.
Section \ref{yuk}  discusses  the contraints from
precision flavour observables, such as
 $(g-2)_\mu$ and  $\tmg$, and  section
\ref{coll} studies the sensitivity 
 to the $\phi -\tau- \mu$ Yukawa
coupling of 
 $\phi \to \tau \bar{\mu}$
at  hadron colliders, in
the light of the rare decay data. 
%


\section{Notation}
\label{notn}

We consider a 2HDM of Type III, in the classification\footnote{This definition
differs from the classification given in \cite{Barger:1989fj},
where there are 4 types of 2HDM, all  of which 
avoid tree level flavour changing neutral currents.} 
used, for instance, in \cite{Davidson:2005cw}.
Contrary to ``type I'' and ``type II'' models,   the
type III scalar potential has no 
discrete symmetry that distinguishes doublets.
When Yukawa couplings are included,  
tree-level flavour changing couplings
for the neutral Higgses are possible. In an arbitrary choice of
basis in doublet Higgs space, the potential can be written \cite{Gunion:2002zf}
\beqa  \label{pot}
\mathcal{V}&=& m_{11}^2\Phi_1^\dagger\Phi_1+m_{22}^2\Phi_2^\dagger\Phi_2
-[m_{12}^2\Phi_1^\dagger\Phi_2+{\rm h.c.}]\nonumber\\[6pt]
&&\quad +\half\lambda_1(\Phi_1^\dagger\Phi_1)^2
+\half\lambda_2(\Phi_2^\dagger\Phi_2)^2
+\lambda_3(\Phi_1^\dagger\Phi_1)(\Phi_2^\dagger\Phi_2)
+\lambda_4(\Phi_1^\dagger\Phi_2)(\Phi_2^\dagger\Phi_1)
\nonumber\\[6pt]
&&\quad +\left\{\half\lambda_5(\Phi_1^\dagger\Phi_2)^2
+\big[\lambda_6(\Phi_1^\dagger\Phi_1)
+\lambda_7(\Phi_2^\dagger\Phi_2)\big]
\Phi_1^\dagger\Phi_2+{\rm h.c.}\right\}\,,
\eeqa
where $m_{11}^2$, $m_{22}^2$, and $\lam_1,\cdots,\lam_4$ are real parameters.
In general, $m_{12}^2$, $\lambda_5$,
$\lambda_6$ and $\lambda_7$ are complex, but  we neglect
CP violation in this paper for simplicity,
and take them  $\in \Re$.  A clear discussion
 of CP violation  can be found in \cite{Branco:1999fs,Accomando:2006ga}.
A translation dictionary to the form of potential  used, 
for instance, in \cite{HHG}, can be found in
\cite{Gunion:2002zf}.

The scalar fields will
develop non-zero vacuum expectation values (vevs) if the mass matrix
$m_{ij}^2$ has at least one negative eigenvalue.
Then, the scalar field
vacuum expectations values are of the form
\beq \label{potmin}
\langle \Phi_1 \rangle={1\over\sqrt{2}} \left(
\begin{array}{c} 0\\ v_1\end{array}\right), \qquad \langle
\Phi_2\rangle=
{1\over\sqrt{2}}\left(\begin{array}{c}0\\ v_2\,
\end{array}\right)\,,
\eeq
where $v_1$ and $v_2$ are real and non-negative,   and
\beq \label{v246}
v^2\equiv v_1^2+v_2^2={4\mw^2\over g^2}=(246~{\rm GeV})^2 = 
\frac{1}{\sqrt{2} G_F}\,
~~~~\tan \b \equiv \frac{v_2}{v_1}
\eeq


The scalar potential of the  type III 2HDM, with real  couplings,
has  three parameters in $m_{ij}^2$, 
four in $\lambda_{1..4}$ and  three in $\lambda_{5..7}$. One
of these parameters can be set to zero by
a basis choice  (for instance $m_{12}^2 = 0$),
leaving 
 nine independent  parameters.
The minimisation conditions  give  $v$, which is measured,
so 8 inputs are required.
Ideally, one would like to express these parameters
in terms of observables, such as masses and 3 or 4 point
functions. This would avoid confusion stemming from the arbitrary
basis choice in Higgs space, and clarifies the measure
on parameter space to use in numerical scans.

However, to  study
the current bounds on  ``light''  Higgses, we only
need  their masses and couplings to SM particles
(controlled by $(\b - \a)$ --- see the discussion
after eqn (\ref{scalareigenstates})). So we
do not need  a complete  parametrisation in terms of
observables; we relate constraints on the scalar potential
to the masses in the basis independent notation
of  \cite{Davidson:2005cw}, and otherwise  our parameters are
 $m_{h}^2,m_{H}^2, m_A^2$,  $m_{H+}^2$,  $\sin (\b - \a)$
and the flavour changing  Yukawa coupling   that
controls $p\bar{p} \to \phi \to \tau 
\bar{\mu}$ and $\tau \to \mu \gamma$.

\subsection{Basis choice in Higgs space and $\tan \beta_\tau$}
\label{sectb}

It is always possible to  choose a basis in  the Higgs  space, 
such that only one doublet acquires a vev \cite{Branco:1999fs}. 
This is known as the ``Higgs basis'', defined such that  
$\langle H_1 \rangle \neq 0$, 
and  $\langle H_2\rangle =0$, and in this basis  
all the potential parameters are written in upper case:
\bea
\mathcal{V}&=& M_{11}^2H_1^\dagger H_1+M_{22}^2 H_2^\dagger H_2
-[M_{12}^2 H_1^\dagger H_2+{\rm h.c.}]~~~~~~~~~~~~~~ {\rm (in~ Higgs~ basis)}
\nonumber\\[6pt]
&&\quad +\half\Lambda_1(H_1^\dagger H_1)^2
+\half\Lambda_2(H_2^\dagger H_2)^2
+\Lambda_3(H_1^\dagger H_1)(H_2^\dagger H_2)
+\Lambda_4(H_1^\dagger H_2)(H_2^\dagger H_1)
\nonumber\\[6pt]
&&\quad +\left\{\half\Lambda_5(H_1^\dagger H_2)^2
+\big[\Lambda_6(H_1^\dagger H_1)
+\Lambda_7(H_2^\dagger H_2)\big]
H_1^\dagger H_2+{\rm h.c.}\right\}\,,
\eeqa
This
follows the notation of \cite{Davidson:2005cw}.
  
In the Higgs basis,  the   angle $\alpha_{HB}$  rotates to
the  mass
basis  of the CP even Higgses $h,H$:
\bea
\hl &=&-(\sqrt{2}~{\rm Re\,}H_1^0-v)\sin \alpha_{HB}+
(\sqrt{2}~{\rm Re\,}H_2^0) \cos \alpha_{HB} \,,\nonumber\\
\hh &=&\phm(\sqrt{2}~{\rm Re\,}H_1^0-v)\cos \alpha_{HB}+
(\sqrt{2~}{\rm Re\,} H_2^0)\sin \alpha_{HB}\,.
\label{scalareigenstates}
\eea
In a 2HDM of  type I or II, there is also a choice
of basis where the discrete symmetry  $\Phi_i \leftrightarrow \Phi_i$,
or  $\Phi_j \leftrightarrow -\Phi_j$ is manifest. Usually,
the lagrangian is written in the basis where this
symmetry is manifest \footnote{For a discussion
of Lagrangian basis dependence in the 2HDM, and
a formalism that is ``basis independent'', see,
{\it e.g.} \cite{Davidson:2005cw,Ivanov:2005hg,Accomando:2006ga}.}, 
the angle $\b$ is 
defined between this ``symmetry eigenbasis''
and the Higgs basis, and  the angle $\a$  rotates between the
symmetry basis and the CP-even mass basis. In which
case
\beq
\alpha_{HB} = \b - \a~
\label{ahb}
\eeq
 and we shall write it 
as such in this paper. Then
the Higgs-$W^+ W^-$ couplings are $ i g m_W C_{\phi WW} 
 g^{\mu \nu}$
with
\beq
 C_{h WW} = s_{ \b - \a} ~,~ C_{H WW} = c_{\b - \a} ~,~  
C_{A WW} =0
\label{phiWW}
\eeq
where  
\beq
 s_{ \b - \a}   \equiv \sin ( \b - \a)~,~
  c_{ \b - \a}  \equiv \cos( \b - \a) ~.
\label{sbadfn}
\eeq
The trilinear couplings between a neutral Higgs and a pair of charged Higgses
are $- i v C_{\phi H+ H-} $, with \cite{Gunion:2002zf}
\beq
 C_{h H+H-} = \Lambda_3 s_{ \b - \a} - \Lambda_7 c_{\b-\a} ~,~
 C_{H H+H-} = \Lambda_3 c_{ \b - \a} + \Lambda_7 s_{\b-\a} ~,~
 C_{A H+H-} = 0
\label{phiH+H-}
\eeq

 In  a type III
model, there is no ``symmetry basis'',
so  $\tan \b$ can not be defined from
the scalar potential. To obtain
a type III potential, it is not sufficient to
write down a Lagrangian with $\lambda_6,  \lambda_7 \neq 0$;
one must also check that it is not
a type II or type I model, in a basis
rotated with respect to the symmetry basis.
For this there are  basis independent
``invariants'', discussed for instance
in \cite{Davidson:2005cw},  which vanish in the presence
of symmetries. 

In phenomenological calculations,  $\tan \b$ usually 
appears in  Yukawa interactions, where it
parametrises the relative size of the Yukawa
couplings and the fermion masses. Whether
it can be defined from the scalar potential
of the Higgses is secondary.  Various ``definitions''
of $\tan \b$, involving fermion masses in the
Higgs basis,  can be envisaged 
\cite{Freitas:2002um,Davidson:2005cw,Baro:2008bg,Haber:2006ue}.  
To maintain the intuition of $\tan \beta$ as
the  relative size of the $\tau$  Yukawa
coupling and $\sqrt{2} m_\tau/v$, we  define 
$H_\tau$ to be the Higgs that  couples to the $\tau$,
and   $\b_\tau$ as the
angle in Higgs doublet space 
between  $H_1$ (the vev) and $H_\tau$:
\bea
H_u &  = & \widetilde{H}_1 \sin \b_\tau +
\widetilde{H}_2 \cos \b_\tau \nonumber  \\ 
H_\tau  &  = & {H}_1 \cos \b_\tau - 
{H}_2 \sin \b_\tau 
\eea 
We choose the $\tau$ Yukawa, because we are
interested in $\tau$ flavour violation. 
This reduces to the usual definition in a
Type II (SUSY) model, where 
 $\widetilde{H} = i \sigma_2 H^*$.

Finally, notice that we  pursue a ``bottom-up''  approach, where we
treat $\tan \beta_\tau$ as a physical parameter, and calculate
processes which are finite. This allows us to neglect 
issues related to the renormalisation of $\tan \beta_\tau$
\cite{Freitas:2002um,Beneke:2008wj,Gorbahn:2009pp},
such as its numerical stability, and additional factors of
$\tan \beta_\tau$ that may appear. 

\subsection{The leptonic Yukawa couplings}
\label{lepY}

In  the Higgs basis for the Higgses, and the mass
eigenstate basis for the $\{ u_R, d_R,e_R, d_L, e_L \}$, the
Yukawa interactions of a type III 2HDM can be written
\bea \label{ymodel2}
\!\!\!\!\!\!\!\!
-{\cal L}_{\rm Y}&=&  \sqrt{2} {\Big (}
\overline{q_L}_{j}  \widetilde{H}_1 \frac{K^*_{ij} m^{U}_{i}}{v} u_{Ri}   +
\overline{ q_L}_i H_1  \frac{m^{D}_{i}}{v}  d_{Ri}
+\overline{ \ell_L}_i H_1  \frac{m^{E}_{i}}{v}  e_{Ri} {\Big ) }
\nonumber \\
&&
+\overline{q_L}_{i}  \widetilde{H}_2 [\rho^{U}]_{ij} u_{Rj}   +
\overline{ q_L}_i H_2 [\rho^{D}]_{ij} d_{Rj}
+\overline{ \ell_L}_i H_2 [\rho^{E}]_{ij} e_{Rj}
+{\rm h.c.}\,,
\eea
where $K$ is the CKM matrix, 
  $\overline {\ell} H_1 = \bar{\nu} H^+_1 + \bar{e} H^0_1$,
 $\widetilde{H}_i = i \sigma_2  H_i^*$, and 
 the generation indices are written explicitly. 
We are principally interested in the leptons, so we drop the 
superscript of $\rho^E \to \rho$. 

If  the  neutral CP-even Higgses 
are defined to be $h$ and $H$, with $m_h \leq m_H$, 
if $A$  is  the CP-odd Higgs,  and 
if flavour violating neutral couplings 
are allowed in the lepton sector only, this gives
the following Yukawa couplings of
leptons in the Higgs mass basis
\beqa \label{FVcplings}
-\call_Y^{leptons} &=&
\overline{ e}_i \left[{m_{i} \over v} \delta_{ij}  s_{ \b - \a} 
+\frac{1}{\sqrt{2}}
([\rho]_{ij} P_R+[{\rho}^\dagger]_{ij} P_L) 
 c_{ \b - \a}
\right]e_j \hl
\nonumber \\[4pt]
&&+ \anti e_i \left[ { m_{i} \over v} \delta_{ij}  c_{ \b - \a}
-\frac{1}{\sqrt{2}}
([\rho]_{ij} P_R+[{\rho}^\dagger]_{ij} P_L)
 s_{ \b - \a} 
\right]e_i \hh
\nonumber\\[4pt]
&&
+\frac{i}{\sqrt{2}}\anti e_i ([\rho]_{ij} P_R-[{\rho}^\dagger]_{ij} P_L)e
\ha\nonumber\\[4pt]
&&
+ {\Big \{ }
 \bar{u}_i \left[[K \rho^D]_{ij} P_R- [\rho^{U \dagger} K]_{ij} P_L\right] d 
H^+ +
  \anti \nu_i [ U^\dagger \rho]_{ij} P_R  e H^+
+{\rm h.c.}  {\Big \} }
\eeqa where we included the charged Higgs interactions
with the quarks.
$U$ is the PMNS matrix,  which we henceforth drop,
assuming that  the neutrinos are in the ``flavour'' basis.
For  the quarks with flavour diagonal Yukawa couplings 
the Lagrangian can be obtained by taking  $[\rho]$ diagonal,
and  substituting $\{ \rho^E, e \} \to \{\rho^D, d\}$
for down quarks, and $\{ \rho^E, e \} \to \{\rho^U, u\}$
for the ups. 

From  eqn (\ref{FVcplings}),
The  neutral Higgs couplings to fermions are :
\bea
g_{h,f f'}& = &
\frac{g m_f}{2 m_W}  s_{ \b - \a}  \delta_{f,f'} + 
\frac{\rho_{ff'}}{\sqrt{2}}   c_{ \b - \a} \nonumber  \\
g_{H,f f'}& = &
\frac{g m_f}{2 m_W}  c_{ \b - \a} \delta_{f,f'}
- \frac{\rho_{ff'}}{\sqrt{2} }  s_{ \b - \a}  \nonumber \\
g_{A,f f'}& = & i \frac{\rho_{ff'}}{\sqrt{2}}  ~~~,  
\label{yf}
\eea
where $g_{A,f f'}$ appears in the Feynman rule with
a $\gamma_5$. 
For simplicity, we will take $[\rho^E]$ hermitian,  and
normalise it 
\beq
[\rho]_{ij} =  - \tilde{\kappa}_{ij}\sqrt{\frac{2m_i m_j}{v^2}} 
 = - \kappa_{ij} ~\tan \b_\tau  \sqrt{\frac{2m_i m_j}{v^2}}
\label{norm}
\eeq
It is tempting to expect  
$|\tilde{\kappa}_{ij}| \sim 1$ \cite{Cheng:1987rs}.
However, recall the definition of $\b_\tau$ from  the end of
 section \ref{sectb}: in the Higgs basis
for the scalar  doublets, and the mass eigenstate basis
of the charged leptons,  the $\tau$  Yukawa coupling is 
\beq
 \overline{\ell_L}\, [ \frac{ \sqrt{2} m_\tau}{v}\, H_1 
+ \rho_{\tau \tau}\, H_2 ]\, \tau_R + h.c.
 \equiv y_\tau \overline{\ell_L}\, [ \cos \b _\tau \, H_1 - \sin \b _\tau 
\, H_2 ]\, \tau_R + h.c.
\label{defntb}
\eeq 
This gives  $\tilde{\kappa}_{\tau \tau } = \tan \beta_\tau $ 
so we factor this out 
and ``expect'' that $\kappa_{ij} \sim 1$. Recall that
$\alpha_{HB} = (\b - \a)$ is a physical mixing angle defined
from the scalar potential, not the difference
of two angles. In particular, the $\b$ in
$(\b - \a)$ is unrelated  to  $\tan \b_\tau$.

We are  interested in
$\tau - \mu$ lepton flavour violation,
so we allow an arbitrary $\kappa_{\tau \mu}$,
and assume that  $\kappa_{\tau e} \sim  \kappa_{e \mu} 
\sim 0$.  To perform a general analysis, we should
treat the $\rho_{\phi tt}$ and  $\rho_{\phi bb}$ as free
parameters, because the angle $\beta_f$, defined for a
fermion $f$ in anology with eqn (\ref{defntb}),
could be different for each $f$.  However,
 we attribute Type II values:
\beq
 [\rho^D]_{ij}  =   -\sqrt{2}   \tan \b_\tau ~ \frac{m^D_{i}}{v} ~ \delta_{ij} ~~,~~~~~~
  [\rho^U]_{ij}   =  \sqrt{2}   \cot \b_\tau~ \frac{K^\dagger m^U_{i}}{v}~ \delta_{ij} ~~,~~~~~~~~~~
  [\rho^E]_{\tau \tau}  = - \sqrt{2}   \tan \b_\tau ~\frac{m_{\tau}}{v}   
~~~~~. 
 \label{ktb}
\eeq
 to all   elements of the  $[\rho]$ matrices 
(except $\rho_{\tau \mu}$), because this parametrisation is
adequately representative of the cases we are interested in
(see the discussion at the end of section \ref{sec:tmg}). 
The expressions of eqn (\ref{ktb})
apply in the mass eigenstate bases of $\{ d_L, e_L, d_R, u_R,e_R \}$. 
For a Type III model  which has  Type II
couplings plus small corrections (as  can arise
in Supersymmetry\cite{Antaramian:1992ya,Brignole:2003iv}),
 the $\rho-$coupling to the top quark is suppressed at
large $\tan \beta$.
 If in addition, either $s_{ \b - \a}$ or $ c_{ \b - \a}$ is small,
eqn (\ref{yf}) shows that 
the CP-even Higgs with larger flavour -violating coupling, will
be weakly coupled to the top.  This suppresses both
the diagrams in figure \ref{bz}.

There are many notations for Type III Yukawa couplings.
Our $\kappa$ bears no relation
to the one in \cite{Davidson:2005cw},
but is proportional to the  $\kappa$ of 
\cite{Kanemura:2005hr}(KOT), who define:
\beq
\frac{m_\tau}{v \cos^2 \b} (\kappa^L_{\tau \mu}   
+ \kappa^R_{\mu \tau} ) {\Big |}_{KOT}
 =  - \kappa_{\tau \mu} \tan \b_\tau \sqrt{ \frac{m_\tau m_\mu}{v^2 }} 
 {\Big |}_{this~paper}
\eeq
The additional power of $1/ \cos \beta$ in the LFV Lagrangian  of KOT
causes their LFV rates to scale as $\tan^6 \b$, rather than
$\tan^4 \b_\tau$ as we  find. Such differences must be taken into
account in comparing plots. To ensure that our results are as
``physical'' as possible, we plot bounds on $(\kappa_{\tau \mu} \tan \b_\tau)$,
as a function of Higgs masses, $ s_{ \b - \a} $ from eqns
(\ref{ahb},\ref{sbadfn}) and $\tan \b_\tau$
defined from eqn (\ref{defntb}).


\section{Higgs mass bounds} 
\label{bds}

In this section, we list bounds on   the parameters
of the scalar potential of eqn (\ref{pot}). These bounds
were recently presented  in basis-independent notation, 
allowing for CP-violation, in \cite{O'Neil:2009nr},
where references to the earlier literature can also be found. 
The bounds can be divided into two classes:
firstly, those which apply directly to the physical masses
and coupling constants.  These  
arise from precision electroweak  analyses
(the $T$ parameter), and are the most stringent.
Secondly, there are bounds on the $\lambda_i$ couplings,
which follow from   imposing  that
$WW \to WW$ is unitary at tree level, and from various
considerations about the Higgs potential (positive,
perturbative...). These must be re-expressed as
bounds on the masses.  In our ``phenomenological'' approach,
where we allow arbitrary New Physics at the TeV scale,
these bounds are less important. We review them briefly
anyway.

\subsection{ Bounds on  the  potential}

We are interested in bounds on the various Higgs masses.
In the  Higgs basis, these are related to potential parameters
as \cite{Davidson:2005cw} :
\bea
\label{ml1}
m_{H+}^2 & = & M^2_{22} + \frac{v^2}{2} \Lambda_3 \\
m_A^2 - m_{H+}^2&= & -\frac{v^2}{2} (\Lambda_5 - \Lambda_4) \\
m_H^2 + m_h^2-  m_A^2 & = & + v^2 (\Lambda_1 + \Lambda_5) \\ 
(m_H^2 - m_h^2)^2 & = & 
[m_A^2 + (\Lambda_5 - \Lambda_1)v^2]^2 + 4 \Lambda_6^2 v^4  \\
\sin[ 2(\b - \a)] & = &  -\frac{2 \Lambda_6 v^2}{m_H^2 - m_h^2 }
\label{ml5}
\eea
The  $H^\pm$ mass can be raised
high enough to respect $b$ physics bounds (see section \ref{flav})
by increasing   $M_{22}^2$  (the decoupling limit \cite{Gunion:2002zf}). 
To keep  $H,A$ light in this limit  requires large $\Lambda$s.

\subsubsection{  vacuum stability/bounded from below (lower bound on $m_\phi$) 
}

The requirement that the electroweak vacuum be stable,
or that the Higgs potential be bounded from below, gives a lower
bound on $m_h$. See \cite{Wells,Sher:1988mj} for a review. The constraint  
is imposed at tree level; one can also include radiative corrections
and  check  that  potential remains
bounded from below. 

Neccessary and sufficient
conditions to obtain  $V(v_i \to \infty) >0$, in the 
softly broken 2HDM type II (in the basis
where $\lambda_6 = \lambda_7 = 0$)
were given in  \cite{Gunion:2002zf}.
A basis-independent analytic discussion of
Type II and  Type III can be found in \cite{ILambdabds}.
However, the Type III bounds do not give simple
analytic formulae;  a more extensive
and numerical analysis was performed by  \cite{FJ,FSB}, 
who find a lower bound
\beq
m_H >121 ~{\rm GeV ~~.}
\eeq

\subsubsection{  triviality/perturbativity (upper bound on $m_\phi$) }

An upper bound on the Higgs masses (which are
$\propto  \lambda_i$) can be obtained from
requiring that the  $\lambda_i$ couplings remain
perturbative at higher energy  scales. One can,
for instance,   impose that the Landau Pole
of the Higgs couplings be at some sufficiently
high scale. The analysis of \cite{FJ}
finds no  upper bound on $m_{H,A,H+}^2$
for a softly broken Type II, or Type III.

\subsubsection{Unitarity}

In the spontaneously broken electroweak
theory, there are
delicate cancellations in
the tree-level amplitude  
for longitudinal $W$ scattering,
 between  diagrams with Higgs or
gauge boson exchange. 
In addition, at scales $\sqrt{s} \gg $ the Higgs masses,
the  various 
scattering amplitudes are  proportional 
to combinations of the $\lambda_i$  couplings.
S-matrix unitarity gives an upper bound on
the S-matrix elements. If the tree
level calculation  is a good approximation
to the full S-matrix,  then this bound  translates
to  
bounds on the Higgs masses. 
These constraints are most interesting for the
$J = 0$ partial wave.  This is discussed 
for the Standard Model in \cite{HHG,LQT},
and  in \cite{KKT,Ginzburg:2005dt} for the 2HDM
(see also the appendix of \cite{O'Neil:2009nr}).

S-matrix unitarity implies that
$1 = S S^\dagger$$ = (1 + i T)(1 - i T^\dagger)$.
Writing the S-matrix between 
states labelled by angular momentum $J, K,..$ 
and all other quantum numbers
called $j,k...$, this gives
\beq
- 2 {\rm Im} \{ \langle j, J | T| k, K \rangle \}
= \langle j, J | TT^\dagger| k, K \rangle
\eeq
which, applied  to the  
 $J = 0$ component of an amplitude:
 $$ a_0 = \frac{1}{16 \pi s} \int_{-s}^0 dt {\cal A}
$$
implies $  2 | {\rm Im} \{ a_0 \}| > |a_0|^2$.

There are two limits in which unitarity bounds are
usually calculated.  The standard
calculation assumes $m_W^2 \ll m_\phi^2 \ll s$, 
neglects all masses($\phi$ is an
arbitrary Higgs), describes the longitudinal gauge bosons
as Higgses via the Equivalence Theorem(see
\cite{Veltman:1989ud} for a pedagogical discussion), and
obtains bounds on linear combinations of the
$\Lambda_i$  (see  \cite{Ginzburg:2005dt,O'Neil:2009nr}
for a basis-independent calculation in the 2HDM). 
Notice that in the strict Type II model ($m_{12}^2 = 0$ in the basis where
$\lambda_6 = \lambda_7 = 0$), these bounds are quite
sensitive to $\tan \b$ \cite{KKT};
this may be related to the  appearance of a
massless Higgs,  in the limit
where one of the vevs $v_i \to 0$.  
This is no longer the case
when  $m_{12}^2  \neq 0$ is allowed \cite{Akeroyd:2000wc,Ginzburg:2005dt}.

Here, we are more interested in the
limit  $m_W^2 \ll s \ll  m_\phi^2$, where $\phi$ 
is some of  the Higgses. That is, we allow
arbitrary New Physics at the TeV scale to
preserve S-matrix unitarity in the
 $m_W^2 \ll m_h^2 \ll s$ limit, but we would
like to know the bounds arising on mass splitting
among the Higgses, at $\sqrt{s} <$ TeV, when
some Higgs are heavier than $\sqrt{s}$, and some
are lighter.

The gauge contribution to longitudinal
$W$ scattering (no Higgses  exchanged in $s$ or
$t$ channel), is \cite{LQT}
\beq
{\cal A} (W_L W_L \to W_L W_L)
= \frac{ig^2}{4m_W^2} (s+t) + ... ~~~~~~~~~~~~~{\rm no~Higgs}.
\label{mine}
\eeq
Requiring $a_0< 1$ implies  the scale
of Higgs masses  should be $\lsim $ TeV.
In the case where some Higgses are light, 
for instance $m_h <s$, including them
in the amplitude reduces the coefficient of
$s+t$, and raises the upper bound on
the masses of the remaining Higgs.
We conclude that unitarity constraints do not give us
relevant bounds on the mass differences among
the Higgses.

\subsubsection{ Electroweak precision  tests } 
\label{secSTU}

New physics that couples weakly to Standard Model fermions, 
but has electroweak gauge interactions,
can be constrained by  the measured values of the
``oblique parameters'' \cite{PDG,Maksymyk:1993zm}.
If the vacuum polarisation tensor between gauge bosons
$i$ and $j$ is defined as
\beq
\Pi^{\mu \nu}_{ij}(q) = g^{\mu \nu} A_{ij}(q^2)
+ q^\mu q^\nu B_{ij}(q^2)
\label{pi}
\eeq
then the   parameters  $S$ and  $T$
can be defined as \cite{Grimus:2007if}
\bea
\label{STU}
\bar{S} = \frac{\alpha}{4 s_W^2 c_W^2} S  & = &
\frac{A_{ZZ}(m_Z^2) - A_{ZZ}(0)}{m_Z^2 } + 
\left.\frac{\partial A_{\gamma \gamma}}{\partial q^2} \right|_{q^2 = 0}
+ \frac{c_W^2 - s_W^2}{ s_W c_W} 
\left.\frac{\partial A_{\gamma Z}}{\partial q^2} \right|_{q^2 = 0} \\
\bar{T} = \alpha T  & = &
\frac{ A_{WW}(0)}{m_W^2 }
-  \frac{ A_{ZZ}(0)}{m_Z^2 } 
 \eea
where $s_W = \sin \theta_W$. 
The Standard Model contributions to these
parameters, including
that of the SM Higgs,  
are assumed to be subtracted out.  
The contributions to $S,T,$ and $U$, (as well as $V,W$ and $X$),  
due to an arbitrary number of Higgs doublets and singlets
have recently been calculated in \cite{Grimus:2007if}.
$S$, $T$ and $U$ were calculated in the  CP violating  2HDM
in   \cite{O'Neil:2009nr}.  We use here the formulae
of \cite{Grimus:2007if}, which verify the previous
calculations of \cite{Chankowski:1999ta,Bertolini:1985ia}
for the 2HDM.

As discussed in  \cite{O'Neil:2009nr}, the 2HDM contributions
to $S$ and $U$ tend to be small enough, but the contribution
to $T$ can exceed  the value allowed for New Physics
( $- 0.15 < T < 0.20$ at one $\sigma$ \cite{PDG}). 
We impose the bound  $-0.05 < T-S < 0.10$ at one $\sigma$\cite{PDG}. 
The calculation
of $T$ in multi-Higgs models is presented in detail
in  the first paper of  \cite{Grimus:2007if}. They give
\bea
T &= & \frac{ 1}{16\pi s^2_W  m_W^2}
        \left\{  F(m^2_A,m^2_{H+})  + \cbmaii  [ F(m_{H+}^2,m^2_h)
 -  F(m_A^2,m^2_h)]  \right. \nonumber \\  
&& \left.
+ \sbmaii [ F(m_{H+}^2,m_H^2)  - F(m^2_A,m^2_H)] \right.  \nonumber \\ 
&& \left.
       -3 \cbmaii \left[ F(m^2_Z,m^2_h)  -F(m^2_W,m^2_h) 
 +F(m^2_W,m^2_H) - F(m^2_Z,m^2_H)  \right] \right\} 
\eea
where
\beq
F(x,y) = \frac{x + y}{2} - \frac{xy}{x-y} \ln \frac{x}{y}
\eeq
is a positive function, that vanishes for degenerate masses.
So as masses in the loop split, the contribution increases. 

It is well known, and clear by inspection, that $T$ becomes
small in various limits, 
such 
as $m_A \to m_{H+}$, or  $m_H \to m_{H+}$ when $s_{\b - \a} \to 1$.
When  studying 
 $h$ and $H$ production and decay
at  colliders,  we will assure 
the precision constraint   by imposing
$m_A \simeq m_{H+} \pm 10 $ GeV. 
For $h$ production, we do not consider the parameters 
  $m_H \to m_{H+}$ and   $s_{\b - \a} \to 1$,
because
$ \sigma (gg \to h \to \tau^\pm \mu^\mp) \propto
s_{\b - \a}^2  c_{\b - \a}^2 $. 
 The  collider cross-section
for the  pseudo-scalar
$A$ is not suppressed by $ s_{\b - \a} $
or $  c_{\b - \a}$, so when
we study $ \sigma (gg \to A \to \tau^\pm \mu^\mp)$, 
 we can  ensure the precision constraint
by requiring
$m_H \simeq  m_{H+} \pm 10$ GeV and  $s_{\b - \a} \to 1$.
For $m_A > 100 $ GeV, $T-S$ is within the 1 $\sigma$ allowed
range for $ \pi/3 < \b - \a \leq \pi/2$. 

\subsection{Flavour physics bounds}
\label{flav}

The charged Higgs $H^+$ neccessarily has 
tree level flavour-changing couplings, like the
$W^+$. Its mass is therefore constrained by
various flavour-changing observables, such as $b \to s \gamma$
and $B^+ \to \tau^+ \nu$. 
 The various bounds  have  been  discussed in \cite{KOO},
who find  $m_{H+} > 250 GeV$ for $2 < \tan \beta < 20 $,
and $m_{H+} > 300 GeV$ for  $\tan \beta < 50$. 
This limit is  lower than that of  Misiak 
{\it et al.} from $b \to s \gamma$ \cite{Misiak:2006zs},
which   is  $m_{H+} > 300 $ GeV
(at $2 \sigma$), because of differences in the
procedure of extraction of the bound.

The charged Higgs 
 contributes at tree level  to the decay of pseudoscalar mesons $M$. 
Since the SM $W$-mediated amplitude is helicity suppressed,  
the  additional suppression of the Higgs amplitude  is 
 only a factor $\sim  (m_M/m_{H+})^2$\cite{Hou:1992sy}.  
Charged meson decays, such as   $B^+ \to \tau^+ \nu$,
  therefore constrain the mass and couplings of $H^+$
in type I and type II models.  
In addition, they can constrain the flavour-violating
couplings of type III models.
It was shown  in \cite{Masiero:2005wr}, that
 a precise determination of  $R_K = 
\Gamma(K^- \to e \bar{\nu})/\Gamma(K^- \to \mu \bar{\nu})$,
by the NA62 experiment, 
could be a  sensitive test of   the $\rho_{\tau \mu}$ coupling.

\subsection{Summary}

We retain two constraints from this section:
$m_{H+} \gsim 300$ GeV from B physics (as discussed in
section \ref{flav}), and  the $T$ parameter
 will be small enough  in two cases:
either  $m_A \simeq m_{H+}$( we take numerically
$m_A = m_{H+} \pm 10 $ GeV,
or  $m_H \simeq  m_{H+}$, 
with   $\b - \a \gsim  \pi/3$ to ensure  $c_{\b - \a} \to 0$.


\section{Constraints on flavour changing  Yukawa  couplings to leptons}

\label{yuk}

In this section, we include  the lepton  flavour-changing
Yukawa couplings of the Higgses, see eqn (\ref{FVcplings}). 
We consider mass ranges for the neutral Higgs which
are consistant with the constraints discussed in the 
previous sections, and study the sensitivity of
$(g-2)_\mu$, $\tmg$ and $\tau \to \eta \mu$  to   the
the flavour-changing coupling  $\rho_{\tau \mu}$,
introduced in eqn (\ref{norm}).
%

A systematic study of bounds on a  2HDM (Type II),
with   lepton flavour violating  Yukawa couplings, has been performed
in \cite{Kanemura:2005hr}. They
impose constraints on $\rho_{\tau \mu}$ arising from tree and
one-loop contributions to rare decays, then  
study the   $\phi \rightarrow \tau \bar{ \mu}$ decay rate
at colliders.  As this is a Type II analysis,
$\tan \beta$ appears in the formulae. 
In the context of supersymmetric models \cite{BR}, 
it was shown in \cite{Paradisi:2005tk}
that the  bounds from one and
two loop contributions to
$\tau \to \ell \gamma$ 
are more stringent than
those from $\tau  \to 3 \ell$ (except
if the Higgses are very degenerate). 

The anomalous magnetic moment of the muon
can also be sensitive to a  2HDM with 
flavour-violating Yukawa couplings \cite{Assamagan:2002kf,Arcelli:2004af,Nie:1998dg,KL}.
For $\rho_{\tau \mu} \sim 10 \sqrt{m_\tau m_\mu}/v$, 
the 2HDM can fit the current discrepancy  in
$(g - 2)_\mu$. However, as we will show,  the  constraint
from     $\tau  \to \mu \gamma$
precludes the explanation of the  $(g-2)_\mu$
discrepancy,
for large areas of parameter space.

\subsection{The  decay $\tau \to \eta \mu$  }

It was shown  in \cite{Kanemura:2005hr}, 
that $\tau \to \eta \mu$ gives a relevant
constraint on  $\kappa_{\tau \mu}$ for
light  pseudoscalars $A$.  We briefly  review here their discussion,
using the updated bound
\cite{Miyazaki:2007jp}
 $BR (\tau \to \eta \mu) < 6.5 \times 10^{-8}$.

The  branching ratio is
\beq
\frac{BR(\tau \to \eta \mu)}{BR(\tau \to \mu \nu \bar{\nu})} 
\simeq  54 \pi^2  \left(\frac{m_\eta}{m_A}\right)^4 
\frac{f_\eta^2}{m_\tau^2}
|\kappa_{\tau \mu}|^2 \tan ^2 \beta_\tau 
\frac{ m_\mu (m_u \cot \beta_{\tau} + ( m_d -2 m_s) \tan \beta_\tau )^2}{(m_u + m_d + 4 m_s)^2}
\label{tauetamu}
\eeq
where $BR(\tau \to \mu \nu \bar{\nu}) = .17$, and  $f_\eta \simeq f_\pi$
from \cite{Black:2002wh}. The current experimental upper
bound  gives
\beq
|\kappa_{\tau \mu}| \tan ^2 \beta_{\tau}  <  70 \left( \frac{m_A}{100 GeV}\right)^2
\label{tme}
\eeq
where we approximate the final fraction of (\ref{tauetamu}) as $\tan^2 \beta_{\tau}/4$.
The bound (\ref{tme}) agrees with the result given in  \cite{Kanemura:2005hr}.
This is comparable to the bound from
$\tmg$ (see figure \ref{ratio;c2v2}); we see that a light 
$A$ is allowed for small $\tan \beta_{\tau}$.

\subsection{The   dipole effective  operator}
\label{efop}

Bounds on a large class of effective operators
that change $\tau -\mu$ flavour are presented in \cite{Black:2002wh}. 
 Several  bounds arise from the dipole operators,
which can be included in an effective Lagrangian as 
\cite{Raidal:2008jk}
\beq
\frac{C^{ij}}{\Lambda_{NP}^2} \langle H \rangle 
\overline{e_i} \sigma^{\a \b} P_R e_j F_{\a \b} + h.c.
\label{dipop}
\eeq
where
 $i,j$ are the flavours of the external leptons.
The  chirality flipping operator must contain
an odd number of Yukawa couplings, including 
$\rho_{\tau \mu}$ and a flavour diagonal
Yukawa coupling of the external Higgs vev.
The model-dependent coefficient  $C^{ij}/\Lambda_{NP}^2 $,
can be related to a New Physics contribution
$\delta a_\mu$ to the   anomalous
magnetic moment of the muon, 
 and to
the $A_{L,R}$ factors that appear in  
$\tmg$  \cite{Raidal:2008jk}:
\beq
\frac{e \delta a_\mu}{4 m_\mu} = 
\frac{{\bf Re} \{ C^{\mu \mu} \}}{\Lambda_{NP}^2}
\frac{v}{\sqrt{2}} 
~~~~~~~~~~~
 \frac{e m_\tau A^{\tau \mu}_{R}}{2} = 
\frac{C^{\tau \mu}}{\Lambda_{NP}^2} \frac{v}{\sqrt{2}} 
~~~~~~~~~~~
 \frac{e m_\tau A^{\tau \mu }_{L}}{2} = 
\frac{C^{\mu \tau *}}{\Lambda_{NP}^2} \frac{v}{\sqrt{2}} 
\label{reln}
\eeq
The $A_{L,R}$ appear in the  
$\tmg$ branching ratio as:
\beq
\frac{
\Gamma (\tau \to \mu \gamma)}
{\Gamma (\tau \to \mu \nu \bar{\nu})}  = 
\frac{48 \pi^3 \alpha}{G_F^2} \left( A_L^2 + A_R^2 \right)
< 
 \frac{ 4.4 \times 10^{-8} }{.17} 
\label{Gtmg}
\eeq
where we used the current $\tmg$ bound \cite{Aubert:2009tk}
(very similar to  the $BR <4.5 \times 10^{-8}$ of  \cite{Hayasaka:2007vc})
to obtain $ |A_L| = |A_R|  \lsim 10^{-4} G_F$.
  
The current experimental and theoretical
determinations of $(g -2)_\mu$ are\cite{PDG}:
\beq
\frac{(g-2)_\mu}{2} =  a_\mu =
\left \{ \begin{array}{ll} 
11 659 208.0(5.4)(3.3) \times 10^{-10} & expt\\
11 658 471.810(0.016) +691.6(4.6)(3.5)  + 15.4(.2)(.1) 
\times 10^{-10} & QED + hadronic + EW
\end{array}
\right.
\eeq
where the first experimental uncertainty is  statistical
and the second systematic.  The hadronic uncertainties
are from lowest and higher order. The electroweak  contribution,
which at one loop is 
\beq
a_\mu^{EW, 1-loop} \simeq \frac{G_F m_\mu^2}{8 \sqrt{2} \pi^2}\left[\frac{5}{3} + 
\frac{1}{3} (1 - 4 s_W^2) \right]
\label{EWg-2}
\eeq
includes  the   two loop  effects (a $\sim 25 \%$ correction), 
and the electroweak
 uncertainties are from the unknown Higgs mass and quark loop effects.  
The  difference between experiment and theory is
\beq
\Delta a_\mu = a^{exp}_\mu - a_\mu^{SM} = 29.2(6.3)(5.8) \times 10^{-10}
\label{da}
\eeq 
so when we ask our model to ``fit the $(g-2)_\mu$ discrepancy'',
we will ask it to contribute  $\delta a_\mu \sim 
a_\mu^{EW} \sim
+15  \times 10^{-10}$ to the
theoretical calculation of  $ a_\mu$.

We now calculate the bounds from a selection of  
one and two loop diagrams contributing
to   $(g-2)_\mu$ and to $\tmg$. 
We aim  to describe the leading constraints.
The model we consider has many unknown Yukawa interactions, 
so which diagrams give  the most ``interesting'' limits depends on 
what is assumed about the pattern of Yukawa couplings. 
For instance, a two-loop
diagram that is linear in a small Yukawa coupling can give
a more significant bound than a one loop diagram
that is quadratic in that same  Yukawa coupling \cite{Bjorken:1977br}. 
Various theoretically motivated patterns
for the Yukawas have been
discussed in the literature 
\cite{Cheng:1987rs,Antaramian:1992ya,DiazCruz:2004tr},
and many analyses concentrate on the supersymmetric case,
where the flavour-violating Yukawas are loop induced, and
$\rho_{tt} \propto \cot \b$, $\rho_{\tau \mu} \propto \tan \b$.

We only include diagrams  that contain the flavour changing
couplings \footnote{So we neglect the flavour diagonal
contributions to $(g-2)$ --- see the references
at the beginning of section \ref{ad}.} $\rho_{\tau \mu}$.
It appears squared in the lepton flavour conserving
amplitude $a_\mu$, and in the lepton flavour changing
rate $\Gamma(\tau \to \mu \gamma)$. 

At one loop,
three Yukawa couplings are required on the fermion line
of the dipole operator,
so we expect
$C^{\mu \mu} \propto |g_{\phi \tau \mu}|^2 m_\tau$
and 
$C^{\mu \tau} \propto g_{\phi \tau \tau} g_{\phi \tau \mu} m_\tau$.
We do not consider possible large logs  that could arise from
electroweak corrections to these diagrams. 
We do consider two loop contributions to $C^{\tau \mu}$
which have only one Yukawa coupling on the lepton line.
The other end of the $\phi$ propagator must attach somewhere,
and an external Higgs vev is required for the dipole operator. 
If this pair of Higgs couplings can be large
(for instance  gauge, third generations Yukawas, or   scalar self-interactions)
they could compensate the $1/(16 \pi^2)$ relative
to the one-loop diagram.  Therefore we
include the two-loop ``Barr-Zee'' diagrams (figure \ref{bz})
with third generation quarks, and with $W$ bosons.
We  neglect  ``Barr-Zee'' diagrams  with charged Higgses
in the second loop  for simplicity, and because these 
loops are relatively suppressed
\footnote{Such diagrams could  be expected to give significant
contributions, because they  are proportional
to quartic Higgs couplings, 
which can be  $\sim 1$  because we  have mass$^2$
differences  in our Higgs spectrum $\sim v^2$. 
However, the amplitudes are suppressed
by $m_\phi^2/m_{H+}^2$, and there is
a partial cancellation between the $H_+$ loops
with both photons attached in the same place
and with the two photons attached separately
\cite{Chang:1993kw}. The
 $H_+$ loops  are therefore  an order of magnitude smaller
than the top loop, for quartic Higgs couplings $\sim 1$. 
Secondly,  the relevant  Higgs couplings 
(see eqn (\ref{phiH+H-}))
 $\Lambda_3$ and $\Lambda_7$, 
are independent of the Higgs masses (see eqns
(\ref{ml1}) to (\ref{ml5})), so are free parameters
which just add more confusion to the analysis.  
We are reluctant to allow $|\Lambda_3|,
|\Lambda_7|  \gsim 3$, which would
allow  to  (partially) cancel
the top loop.  
}.

\begin{figure}[ht]
  \centerline{
 \scalebox{0.90}{\includegraphics{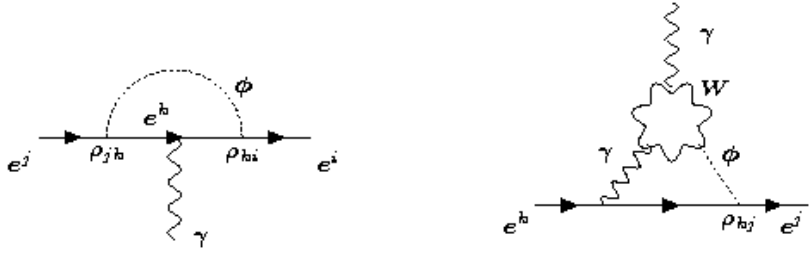}}}
   \caption{\small 
One and  two loop  diagrams, which contribute
  to  the dipole operator coefficient $C^{ij}_{e \gamma}$ 
(see eqn (\ref{dipop})). A mass insertion is required on the fermion
line of the one loop diagram. 
 For  $k = \tau$ and $j= i  = \mu$, 
these diagrams  depend   only on  the Yukawa couplings
$\rho_{\tau \mu}$.}
\label{bzg}
\end{figure}

Our formulae are incomplete, possibly
gauge dependent (the $W$ loop diagrams of figure
\ref{bzg}), and we  have not excluded  large logs 
in higher loop corrections (as in the SM electroweak
contribution to $g-2$ \cite{Czarnecki:1995wq}). 
Due to these uncertainties, we will be
reluctant to exploit cancellations between
diagrams.

\subsection{g-2}
\label{ad}

There are one loop (and higher order) contributions
to $a_\mu$  mediated by the flavour conserving couplings
of the 2HDM. 
The possibility of explaining the  $(g-2)_\mu$ discrepancy 
with these terms   was discussed in \cite{Cheung:2003pw},
who show that there is little parameter space consistent with
other constraints on the 2HDM.  We assume that $(g-2)_\mu$
does not constrain the flavour-conserving parameters of
our model\cite{KOO}.

In the presence of the flavour changing
  $\rho_{\tau \mu}$ coupling, there
is a one-loop contribution to  $(g-2)_\mu$, 
illustrated  on the left in figure \ref{bzg}. As discussed above, 
it should be  of order the one-loop
electroweak gauge contribution, or less. 
The chirality flip is provided  by
the $\tau$ mass, which increases this
diagram with respect to flavour diagonal
ones.
Neglecting the
lepton masses in the kinematics, 
the one loop contribution gives \cite{Assamagan:2002kf}
\bea
 a^{2HDM,LFV}_\mu
 & \simeq & 
\sum_\phi g_{\phi \tau \mu}^{2}  \frac{m_\mu m_\tau}{8 \pi^2} \int_0^1
 dx
\frac{ x^2\, }{ m_{\phi}^2 - x( m_{\phi}^2 - m_\tau^2)}
\nonumber \\
 & \simeq & 
\sum_\phi  g_{\phi \tau \mu}^{2}
 \frac{ m_\mu m_\tau}{8 \pi^2m_{\phi}^2} 
\left(
\ln \frac{ m_{\phi}^2}{  m_\tau^2}
-\frac{3}{2} \right)
\label{Delamu}
\eea
where $\phi = h, H, A$, and $g_{\phi f f'}$ is
from eqn (\ref{yf}).

In the absence of cancellation between
the  opposite sign CP-even and CP-odd Higgs diagrams, 
\beq 
 \kappa_{\tau \mu}  \tan \b_\tau \lsim   
 32 \frac{m_\phi}{100 GeV}
\label{daapp}
\eeq
 contributes at most  the experiment-theory discrepancy in 
$\Delta a_\mu \sim 2 a_\mu^{EW}$ \cite{Assamagan:2002kf}. This bound  
agrees with the naive power counting expectation
that $\frac{m_\tau}{m_\mu}|\rho_{\tau \mu}|^2/m_\phi^2 \lsim G_F$.

\subsection{$\tmg$}
\label{sec:tmg}

The two body decay $\tmg$ can arise via loops, 
and the  current  bound
on the   Branching Ratio is given in eqn (\ref{Gtmg}).
Since  the Standard Model  decay $\tau \to \ell \nu \bar\nu$
 is  three body,  the bound on New Physics in a loop
is particularily restrictive, because the
$|1/16 \pi^2|^2$ loop factor is partially
compensated by the two to three body phase space ratio.

To estimate $BR(\tmg)$ in our model, 
we assume that $A_L = A_R \equiv A$, and neglect the lepton
masses in the kinematics. We include
the one-loop diagram of figure \ref{bzg} (the first sum below), 
and a subset of  two-loop diagrams. Following \cite{Chang:1993kw}, 
we include the two-loop diagram of figure \ref{bz}, with an internal
photon and a third generation quark. This is
the second sum below, with $f$ = $t,b$, and is gauge invariant
on its own. The remainder of  eqn (\ref{WW2loop}) is 
 a subset of the    two loop $W$ diagrams (as in figure \ref{bzg})
 \cite{Chang:1993kw}. The amplitude is:
\bea
  A & \simeq&   \frac{1 }{16 \pi^2} \left(
\sqrt{2} \sum_\phi   
 \frac{ g_{\phi \mu \tau}  g_{\phi \tau \tau}}{m_{\phi}^2}
\left( \ln \frac{m_{\phi}^2}{m_\tau^2} - \frac{3}{2} \right) 
\right. \nonumber \\
& & 
~~~~~ +2  \sum_{\phi,f}   g_{\phi \mu \tau} g_{\phi ff}  
\frac{N_c Q_f^2 \alpha}{ \pi}
\frac{1 }{m_\tau m_{f}} \, f_\phi(\frac{m_f^2}{m_{\phi}^2})    
\label{mtaudenom}  \nonumber\\
&& \left.
-  \sum_{\phi = h,H} { g_{\phi \mu \tau}  C_{\phi WW}}
 \frac{ g \alpha}{2\pi m_\tau  m_W}
\left[ 3 f_\phi(\frac{m_W^2}{m_\phi^2})  + \frac{23}{4} 
g (\frac{m_W^2}{m_\phi^2}) + \frac{3}{4} 
h (\frac{m_W^2}{m_\phi^2})  + m_\phi^2 \frac{f_\phi(\frac{m_W^2}{m_\phi^2})
- g(\frac{m_W^2}{m_\phi^2})}{2m_W^2}
\right]\right)                       \label{WW2loop}
\eea
where $\phi = h, H, A$,  $f$ = $t,b$,  
 the coupling $g_{\phi ff'}$  of the internal loop fermion to the 
scalar $\phi$ is given in 
eqn (\ref{yf}),
and  the scalar-$W^+ W^-$ couplings $C_{\phi WW}$ are given in eqn (\ref{phiWW}).
There is a factor $m_\tau$ in the 
denominator of the two-loop expressions   because it 
appears in the definition (\ref{reln}), and a factor
of the loop mass  because
 the functions
$f(z),g(z),h(z)$ are proportional to this mass$^2$,
whereas the loop has a single  mass insertion.

The various functions are \cite{Chang:1993kw}:
\beq
 f_A(z) \equiv g(z)  =  \frac{z}{2} \int_0^1 dx  \frac{1}{ x(1-x)-z }  \ln \frac{x(1-x)}{z}
~~~~~~~~~{\rm  pseudoscalar}
\label{ps}
\eeq
\beq
 f_{h,H}(z) =   \frac{z}{2}  \int_0^1 \, dx \, 
\frac{ (1-2x(1-x))}{x(1-x)-z}  \ln\frac{x(1-x)}{z}
~~~~~~~~~{\rm  scalars},
\label{sc}
\eeq
and
\beq
 h(z) =    - \frac{z}{2}  \int_0^1  
\frac{ dx}{x(1-x)-z} \left[1 - 
\frac{z}{x(1-x)-z}
 \ln\frac{x(1-x)}{z} \right]
~~~~~~~~.
\label{WW}
\eeq
They are  $\sim z$  for arguments  $z$ of order 1, and for small
 $z$, $f_\phi(z)  \sim \frac{z}{2}( \ln z)^2$.

The $\tmg$ amplitude $A$ depends on
$\tan \b_\tau$,  $(\b - \a)$, and the Higgs masses. 
It is useful to estimate the relative 
size  of the one-loop, two-loop fermion, and
two-loop $W$  contributions in various cases.

\begin{enumerate}

\item
We start by estimating the one-loop
diagrams contributing to  $\tmg$. 
In the absence of cancellations between
the  opposite sign CP-even and CP-odd Higgs diagrams, 
 the approximate $\tmg$ bound $|A| < 10^{-4} G_F$  gives
\beq
\frac{\kappa_{ \mu \tau}  \kappa_{\tau \tau} \tan^2 \b_\tau }{8 \pi^2} 
\sqrt{\frac{m_\mu }{m_\tau}} 
\frac{  m_\tau^2}{m_{\phi}^2} \ln \frac{m_{\phi}^2}{  m_\tau^2}  <
10^{-4}
\label{bd1loop}
\eeq
or, for $\kappa_{\tau \tau} \simeq 1$:
\beq
 \kappa_{ \mu \tau}  \tan^2 \b_\tau \lsim 160 
\left( \frac{100 GeV}{m_\phi} \right) ^2. 
\eeq
which is weaker than    the estimated  $(g-2)$  bound  of eqn (\ref{daapp})
for small $\tan \b_\tau$, and more restrictive as  $\tan \b_\tau$ grows.
In the case $\phi = A$, the bound  (\ref{tme})
from $\tau \to \eta \mu$ is
more restrictive. 

\item Consider now    the
two-loop contributions to $\tmg$, starting with the top loop. For
small $\tan \b_\tau$,  or  large 
mixing
$  s_{ \b - \a} \sim  c_{ \b - \a}$
 between  the Higgses,
this can be the dominant contribution to
$\tmg$. 
The   ratio of the top-loop to  one loop  amplitudes,  
 induced by a particular Higgs $\phi$, is 
\bea
\frac{2 loop ~ top}{1 loop}& \sim& 
\frac{ \alpha  Q_t^2  (c_{ \b - \a} s_{ \b - \a} + \cot \b_\tau) m_t^2}{ m_\tau v m_\phi^2}
\frac{m_\phi^2v}
{ m_\tau \kappa_{\tau \tau} \tan \b_\tau
 \ln (\frac{ m_\phi}{m_\tau} )}
\nonumber \\ &&
 \sim   \alpha  Q_t^2 
\frac{ (c_{ \b - \a} s_{ \b - \a} + \cot \b_\tau) m_t^2}{  m_\tau^2}
\frac{1 }{  \kappa_{\tau \tau} \tan \b_\tau \ln (\frac{ m_\phi}{m_\tau} )}
\label{2l/1l}
\eea
where
$g_{\phi ff '}$ is  from eqn (\ref{yf}).
So for  $\kappa_{\tau \tau} \sim 1$,  and $ c_{ \b - \a} s_{ \b - \a} \sim 1 $,
the 2-loop top diagram dominates
over the one loop for $\tan \beta_\tau \lsim m_t/m_b$.
 This is because
all the Higgses  have an ${\cal O} (m_t/v)$ coupling to
the top, 
independently of $\tan \b_\tau$.
However, the $ c_{ \b - \a} s_{ \b - \a}$ terms of the  $h$ and $H$
amplitudes have opposite sign (recall that $g_{A ff'}$ is independent
of $\b - \a$).   
Imposing the approximate $\tmg$ bound $|A| < 10^{-4} G_F$ 
 on the top amplitude, for  case 1:  $m_h \simeq m_H \ll m_A$,
and  for case 2: $m_h \simeq m_A \ll m_H$ with $s_{\b - a} \sim 1$,
implies
\beq
10  \gsim \left\{
\begin{array}{ll}
\left( 
\frac{m_H^2 - m_h^2}{ m_h^2}\sin 2(\b - \a)
+ \frac{1}{\tan \b_\tau} \right)
 ~\kappa_{\tau \mu} \tan \b_\tau  & {\rm case} 1 \\
  \kappa_{\tau \mu} &   {\rm case} 2 
\end{array}
\right.
\label{tmg;c1}
\eeq
which is somewhat more restrictive that 
the estimated  $(g-2)$ bound of eqn (\ref{daapp}), and one-loop
$\tmg$ bound of eqn (\ref{bd1loop}).  
However,  for large $\tan \b_\tau$,
small $c_{ \b - \a}s_{ \b - \a} < \cot \b_\tau$ 
and  $\kappa_{\tau \tau} \sim 1 $,
 the one-loop contribution is  larger..

\item Superficial inspection suggests that  the two loop diagrams
with an internal  $b$ or $W$ loop,   never provide a dominant
contribution (although they are included in
our plots).
Parametrically, the  two loop contributions of the $t$, $W$ and
$b$  give the following three terms
\beq
\frac{\kappa_{\tau \mu} \tan \b_\tau }{m_\phi^2} \left(
  N_c Q_t^2 m_t^2 [ c_{ \b - \a}s_{ \b - \a} + \cot \b_\tau] +     
c_{ \b - \a}s_{ \b - \a} m_W^2
+ N_c Q_b^2 m_b^2 \tan \b_\tau \ln (\frac{m_\phi}{m_b} ) \right)
\eeq
so one sees that
 the $W$ loop can be neglected
with respect to the tops.
Notice  from eqns (\ref{yf}) and (\ref{phiWW}), that when 
$s_{ \b - \a}$ or $c_{ \b - \a}$
is small, one of $h$ or $H$ is 
``SM-like'', meaning that it has coupling
$\sim g m_W$ to $W^+ W^-$, flavour-diagonal
couplings $\sim m_f/v$ to the fermions, and suppressed
flavour-changing  interactions. So the effective
interaction of $W^+ W^- \tau \bar{\mu}$, induced by
the neutral Higgses, is $\propto \sin 2(\b - \a)$.
Similarly, 
the  Higgs-induced $t \bar{t} \tau \bar{\mu}$ is
$\propto  \sin 2(\b - \a) + {\cal O}(\rho_{\phi tt})$.
 This is in accordance  with 
\cite{Chang:1993kw}, who observe that the $W$ loop
contribution vanishes in the decoupling limit
due to unitarity arguments.  It disagrees with
the MSSM analysis of \cite{Paradisi:2005tk}, where the
$W$ contribution is included but  the tops are not.

Comparing the  one loop amplitude  to the  two loop  bottom amplitude, 
for large $\tan \b_\tau< 1/(c_{ \b - \a}s_{ \b - \a})$  when
the $b$ loop could dominate the top,   one obtains:
\beq
\frac{ 1 loop~}{2loop~bottom}
\sim \frac{ \kappa_{\tau \tau} \tan \b_\tau m_\tau \ln (\frac{m_\phi}{m_\tau})  }
{v m_\phi^2}
\frac{m_\tau v m_\phi^2}
{\alpha Q_b^2 m_b^2 \tan \b_\tau \ln (\frac{m_\phi}{m_b} )}
\sim \frac{\kappa_{\tau \tau} m^2_\tau   }
{\alpha Q_b^2 m_b^2 } 
\eeq
This suggests that unless $\kappa_{\tau \tau} \ll 1$, the one-loop contribution
is larger than the two-loop when $
1/(c_{ \b - \a} s_{ \b - \a}) > \tan  \b_\tau  \gsim 3$.

\item Finally, recall that   cancellations can occur
in the sum. There is a relative
negative sign between the CP-even and CP-odd Higgs
diagrams (which has little effect in many
of our plots because we set $m_A \simeq m_{H+} \simeq 300$ GeV,
to minimise the $T$ parameter),
and also between  the $h$ and $H$ induced
two loop top amplitudes
at negligeable $\cot \b_\tau$, which  
would cancel for  $m_h^2  =  m_H^2$. 


\end{enumerate}

\begin{figure}[ht]
 \centerline{\hspace{-1.5cm}
 \scalebox{0.60}{\includegraphics{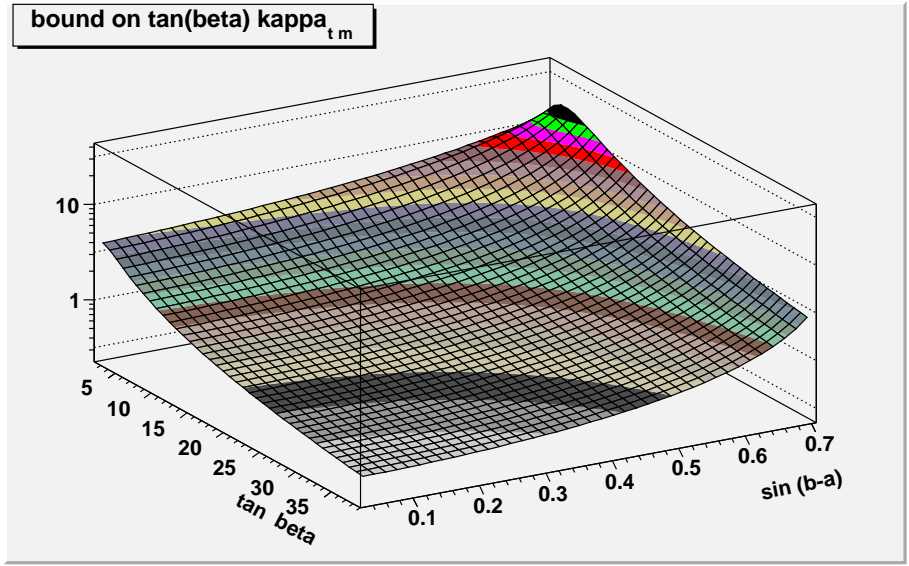}}}
  \caption{\small 
Bound on $ \kappa_{\tau \mu} \tan \beta_\tau$ from the experimental limit on
$BR(\tmg)$, as a function
of $\tan \b_\tau$ and $\sin(\b - \a)$,
 for $m_h = 117$ GeV, $mH = 130$ GeV, $m_{H+} = 300$ GeV
and $|m_A - m_{H+}| < 20 $GeV. 
 We take $\kappa_{\tau \tau} = 1$.
}
  \label{ratio;c2v2}
\end{figure}

The bound on $\kappa_{\tau \mu} \tan \b_\tau$ from
$\tmg$ is plotted in figure \ref{ratio;c2v2},
for $m_h = 117$ GeV, $mH = 130$ GeV, $m_{H+} = 300$ GeV,
$|m_A - m_{H+}| = 10 $ GeV, and  $\kappa_{\tau \tau} = 1$.
As expected, for $\tan \b_\tau \gsim$ few and 
small $\sin(\b - \a)$, the amplitude  is dominated by
the one-loop contribution, and
the constraint on $\kappa_{\tau \mu} \tan \b_\tau$
scales  as $1/\tan \b_\tau$ for fixed $\b - \a$. 

\begin{figure}[ht]
  \centerline{\hspace{1.5cm}
 \scalebox{0.60}{\includegraphics{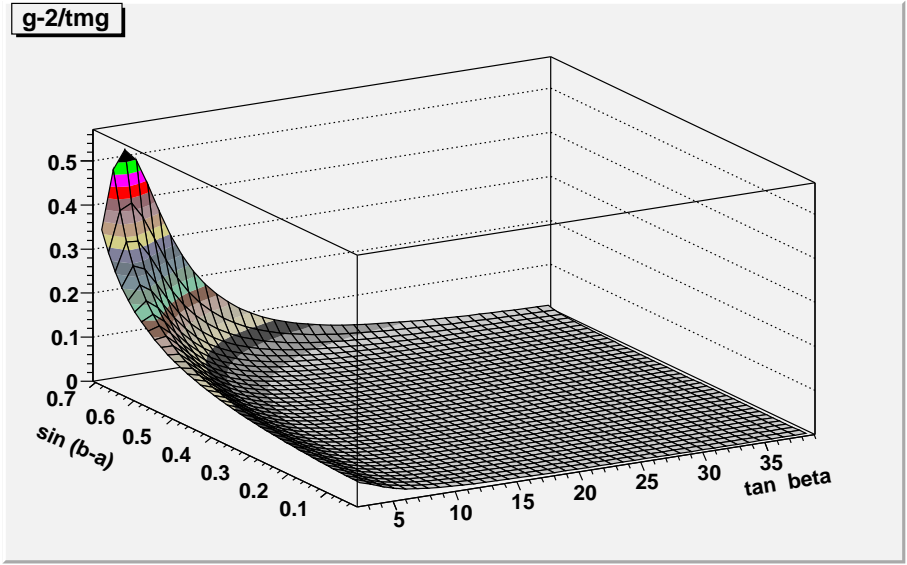}}}
   \caption{\small 
The double ratio,  given in eqn
(\ref{toplot}),
of  the contribution to  $(g-2)_\mu$ divided by
half the experimental discrepancy, over the contribution to
$\tmg$ divided by  its experimental  limit. It is plotted as
a function of $\tan \b_\tau$ and $\sin(\b - \a)$, for  
 $m_h$ = 117 GeV, 
$m_H = 130$ GeV,  $m_{H+} =  300 $ GeV,  and 
$|m_{H+} - m_A|  =  10$
GeV.  Both contributions are $\propto
(\kappa_{\tau \mu} \tan \b_\tau)^2$, which cancels
in the ratio.  Assuming that our
mass choices are representative, 
this shows that the current experimental bound
on $\tmg$  is more restrictive than $(g-2)_\mu$.
}
  \label{ratio;c1}
\end{figure}

The relative importance of the $(g-2)$
and $\tmg$ constraints is illustrated in figure
\ref{ratio;c1}, for a particular choice of
Higgs masses.  The two loop contributions
to $\tmg$ (discussed above) are also included
in this plot. 
 For $\kappa_{\tau \mu} = 1$,
 the double
ratio of the  predicted  $ a^{2HDM,LFV}_\mu$
from eqn  (\ref{Delamu})
over  the $(g-2)$   discrepancy (taken to be $ 15 \times 10^{-10}$),
is divided by the predicted $\tmg$ branching ratio over the current
bound:
\beq
\frac{\frac{a^{2HDM,LFV}_\mu}{\Delta a_\mu}}
{\frac{BR^{2HDM,LFV}(\tmg)}{BR^{data}(\tmg)}}
\label{toplot}
\eeq

Since both  predictions scale as $|\rho_{\tau \mu}|^2$,
this cancels in the ratio, which therefore quantifies
the relative significance of the bounds. 
Since the  ratio $< 1$, we conclude that for generic mass
choices that give a small $T$ parameter, it is not  possible to take
$\rho_{\tau \mu} $ large enough to fit  $(g-2)_\mu$, without
exceeding the bound from $\tmg$.   The plot is obtained
by a grid scan. We consider
$\beta - \alpha : 0 \to \pi/4$, because the range
$  \pi/4 \to \pi/2$ is equivalent if one simultaneously
exchanges \footnote{The  Higgs couplings
are unchanged if one interchanges $h$ with $H$,
and simultaneously   $(\b -\a) \to  (\b -\a) - \pi/2$.
And since  $h$ and $H$ are summed in
the amplitude ${\cal A}$: 
${\cal A}(  c_{ \b - \a} \to 0) =
{\cal A}( s_{ \b - \a} \to 0)  +$ 
corrections due to $m_H^2 \neq  m_h^2$.}  $m_h$ and $m_H$.

\subsection{Assumptions about the $\rho_{ff}$}

As previously mentioned, 
the bounds on $\rho_{\tau \mu}$
depends on the assumptions 
made for the other model parameters, such as
 $\rho_{\tau \tau}$,  $\rho_{tt }$,  $\rho_{bb}$, and 
the Higgs masses. We focus on three limits.

The most conservative  approach to 
 setting bounds on   $\rho_{\tau \mu}$,
is to neglect all other couplings.  In this case,
the $(g-2)$ bound will hold, and vary as a function
of  the neutral Higgs masses, but the $\tmg$
bound plotted in figure \ref{ratio;c2v2} does not apply because
it assumed $\rho$ matrices as given in eqn (\ref{ktb}).
Neglecting  the flavour diagonal Yukawa couplings
of the flavour violating $\phi$,  is
equivalent to setting them to zero. 
So this case  corresponds to
$c_{ \b - \a} = 0$,   where   one 
  CP-even Higgs ($h$) is SM-like, and   
 both $H$ and $A$   have  the flavour-violating   
$\rho_{\tau \mu}$ coupling,
but  no tree-level interactions with $t$s, $b$s or $W$s. 
Such additional  Higgses would be difficult to produce at colliders,
so we do not consider further this case.

We are interested in a Higgs that  could be copiously
 produced at hadron colliders in  $gg$ fusion. This
can be obtained by  allowing  the flavour-violating Higgs
  an ${\cal O} (1) $ Yukawa coupling to the top.
 This  is realised for  $\sin 2(\b - \a) \sim 1$, in the parametrisation 
of Yukawa couplings  given in eqns (\ref{yf}) and (\ref{ktb}).
In this case, the
 top loop contribution to $\tmg$ (see figure \ref{bz})
is significant at small $\tan \b_\tau$,
as discussed in section \ref{sec:tmg}
and the diagram for the $gg \to \tau \bar{\mu}$  process
at colliders  is closely related to the
$\tmg$ diagrams, as expected from the diagrams  in figure \ref{bz}.

 Finally, consider  the large $\tan \b_\tau$
limit. Recall that we defined $\tan \b_\tau$
in our type III model using the tau Yukawa 
coupling (see eqn (\ref{defntb})). 
Then we assumed that all the other
fermions shared this definition of
$\tan \b_\tau$, and had type II Yukawa interactions.
Such  ``almost Type II'' couplings could 
arise in Supersymmetry, where the flavour violating
Yukawas  grow  in the
large $\tan \b$ limit \cite{Brignole:2003iv}.  For large
$\tan \b_\tau$, the bound on $\kappa_{\tau \mu} \tan \b_\tau$
arises from the one loop contributions
to $\tmg$, and is (approximately) independent
of $\sin 2(\b -\a)$. However, the
Higgs-induced $ \bar{t} t \tau \bar{\mu}$ interaction
is $\propto \sin 2 (\b - \a)$, so,  as we will see
in the next section,  
$gg \to \phi \to \tau^\pm \mu^\mp$
is suppressed at small  $ \sin  2(\b - \a)$.
This can be seen by comparing   figures \ref{ratiosigtmg}
and \ref{ratio;c2v2} at small\footnote{ Small $\sin 2 (\b - \a)$ 
can  be obtained   because
either $s_{ \b - \a}$ or  $c_{ \b - \a}$ is small.
Both cases are discussed in the supersymmetric
analysis of  \cite{Paradisi:2005tk}.
Recall that $c_{ \b - \a} \to 0$ in  the decoupling limit
\cite{Gunion:2002zf},  
where $A, H $ and $H^+$ are ``heavy'' ($\gsim 2 m_W$), 
and  the light higgs $h$ has almost SM couplings.}
 $s_{ \b - \a}$.

\subsection{Summary}

 For much of the parameter space of
this model, the  bound from $\tmg$ precludes explaining
the $g-2$ discrepancy (see figure \ref{ratio;c1}). 
We will see in the following section, that 
generic 2HDM  parameters  giving a   detectable rate
for $p \bar{p} \to \phi \to \tau ^\pm \mu^\mp$
in current Tevatron data  
 are already excluded by precision bounds. 
However, a signal at the Tevatron could arise 
if there are cancellations in the 
$\tmg$ amplitude.

\section{Colliders}
\label{coll}

The leading order production cross-section of Higgses,
by gluon-gluon fusion at a $ p \bar{p}$ collider, is \cite{Djouadi:2005gj}
\bea
\label{sigmaLO}
\sigma _{LO} (p\bar{p} \to \phi) & =& 
\frac{ G_F \alpha_s^2 m_\phi^2}{288 \sqrt{2} \pi s} 
\left| \frac{3}{4} \sum_{q = t,b}  g_{\phi q q } {\cal A}  ^\phi(z_q) \right|^2
\int_{z_\phi}^1 \frac{dx}{x} g(x) g(x/z_\phi)
\eea
where $z_\phi = m_\phi^2/s$,  $z_q =(4 m_q^2)/ m^2_\phi$,
 the gluon density in the proton (or anti-proton) is
 $g(x)$, 
\bea
 {\cal A}  ^{h,H}(z) & = &
2z [1 + ( 1 -z) f(z)] \\
 {\cal A}  ^{A}(z) & = &
2 z  f(z)
\eea
and  
\beq
f(z) = \left\{
\begin{array}{cc}
arcsin^2 \sqrt{\frac{1}{z}}  & z \geq 1 \\
-\frac{1}{4}\left[
\log \frac{1 + \sqrt{1 - z }}
{1 - \sqrt{1 - z }} - i \pi \right]^2
& z < 1
\end{array} \right.
\eeq
 The next order QCD corrections
to the production cross section  
are  $\sim 20- 90\%$, and can be  mimicked by
a K factor   \cite{Djouadi:2005gj}.
Finally, to obtain the cross-section for $
p \bar{p} \to \tau \bar{\mu}$, the
cross-section (\ref{sigmaLO}) should be
multiplied by the branching ratio
$BR (\phi \to \tau \bar{\mu})$.

The Tevatron searches for SM Higgses, decaying to
$\tau \bar{\tau}$, in the mass range  105- 145 GeV
\cite{Abazov:2009hm,D0}.
A recent D0 analysis \cite{D0} obtains limits on the
cross-section $\times$ branching ratio of order 
20-90 $\times$ the SM expectation. If we imagine
that the $\tau \bar{\mu}$ final state is  detectable
with similar efficiencies to   $\tau \bar{\tau}$,
then the parameters to which  the Tevatron is sensitive
 can be estimated as follows.
We normalise the $\phi$  production cross-section
and branching ratio 
to the SM expectation for $h_{SM} \to \tau^+ \tau^-$:
\bea
R^\sigma_\phi& \equiv& 
\frac{ \sigma_{LO} (p \bar{p} \to \phi) }
{ \sigma_{LO} (p \bar{p} \to h_{SM}) }
= \frac{ \left| \frac{3}{4} \sum_{q = t,b}  
g_{\phi q q } {\cal A}  ^\phi(z_q) \right|^2 
}{\left| \frac{3}{4} \frac{g m_t}{2 m_W} {\cal A}^h(z_t) \right|^2} 
\label{ratiosigma1}
 \\
R^{BR}_\phi& \equiv &
\frac{  BR  (\phi \to \tau \bar{\mu},\mu \bar{\tau})}
{  BR  (h_{SM} \to \tau \bar{\tau})} \nonumber \\
&\simeq&  \frac{2  
 \kappa^2_{\tau \mu} \tan^2 \b_\tau  m_\mu/m_\tau }
{\left|\frac{v}{m_\tau} g_{\phi \tau \tau} \right|^2(1 -  B_{2W} )  
+  |c_{\phi WW}|^2  B_{2W} } \times
\left \{ 
\begin{array}{cc}
 s_{ \b - \a}^2 &  \phi = h \\
c_{ \b - \a}^2 &  \phi = H \\
1  &  \phi = A
\end{array}
\right. 
\label{ratiosigma}
\eea
where $h_{SM}$ is the Standard Model higgs,     $m_{hSM}$ is taken
to be equal to $ m_\phi$ and  $B_{2W} =   
BR  (h_{SM} \to W^+W^-)$ which varies with $m_h$. 
We estimate that   
\beq
R^\sigma_\phi R^{BR}_\phi  \lsim   30
\label{40}
\eeq
because   the
Tevatron limit on  $\sigma (p \bar{p} \to h_{SM} \to \tau \bar{\tau}) $
is  $\sim 20-90 \times $ the SM expectation \footnote{we 
use   figures from table VI of \cite{D0},
obtained with  $ \sim 5 fb^{-1}$ of data}.
In the large $m_A$ scenario, where the Tevatron would be
looking for a CP-even $\phi$, our lepton flavour violating
rate is maximised for 
  large  $\sin 2(\b - \a) \sim 1$ and small  $\tan \b_\tau = 2$. 
The Tevatron bound would be of order
\beq
 \kappa_{\tau \mu} \tan \b_\tau 
\lsim  \sqrt{ \frac{ m_\tau}{ m_\mu} (20  ~to~ 90)} \sim 20
\label{Tevguess} 
\eeq
which  is on the border of the exclusion
estimate  from $\tmg$, given in eqns(\ref{bd1loop})
and   (\ref{tmg;c1}). 
In the large $m_H$ scenario, where  $\sbma \to 1$,
the Tevatron could look also  for $A \to  \tau^\pm \mu^\mp$.
 At  small $\tan \b_\tau$, the production
rate is similar to the standard model, so
the bound  should be of order eqn (\ref{Tevguess});
for large $\tan \beta_\tau$, $A$ production in $gg$
fusion is suppressed.

\begin{figure}[ht]
 \centerline{\hspace{-1.5cm}
 \scalebox{0.60}{\includegraphics{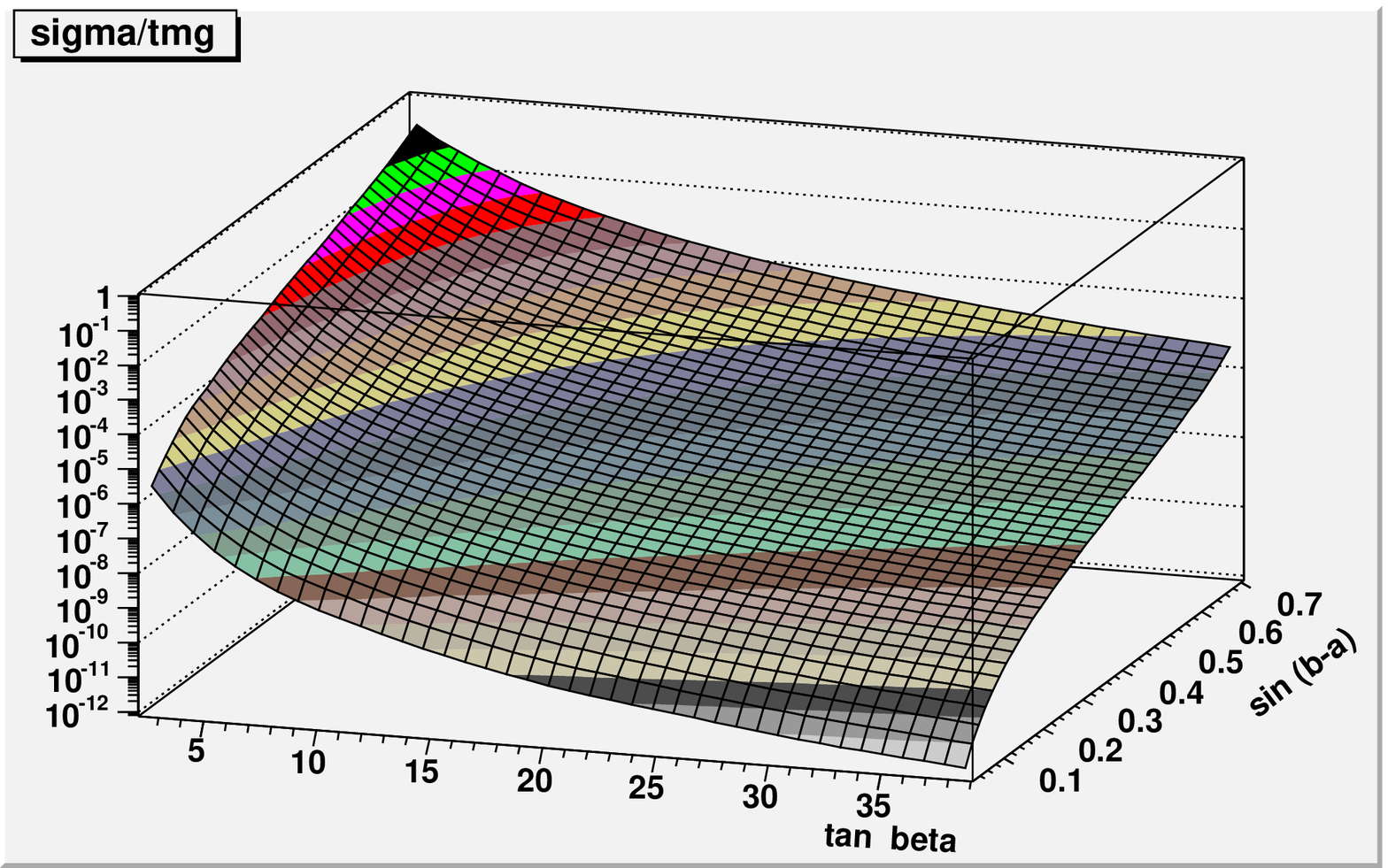}}}
  \caption{\small 
Estimate of the relative sensitivity of
$\sigma(p \bar{p} \to h  \to \tau^\pm \mu^\mp)$
and $\tmg$ to $\rho_{\tau \mu}$ from current data. The plot is
the double ratio of  $R^\sigma_h R^{BR}_h$ (see eqns \ref{ratiosigma}
and \ref{ratiosigma1})
 over $BR(\tmg)$,
divided by current experimental limits (see eqns
(\ref{Gtmg}) and (\ref{40}))
, as a function of
$\tan \b_\tau$ and $\sin (\b - \a)$ ($\kappa_{\tau \mu}
\tan \b_\tau$ cancels in the ratio).  The plot is
for $h$ production at colliders, with
 $m_h = 115$ GeV,  $m_H = 130$ GeV,   $m_{H+} = 300$ GeV
and $|m_A - m_{H+}| = 10 $GeV.  Since the ratio  $<1$, 
$\tmg$ is more sensitive. Alternatively, this plot indicates the
amount of tuning required in the $\tmg$ rate to accomodate 
a signal for $h \to \tau \mu$ at the TeVatron in the near future.
}
  \label{ratiosigtmg}
\end{figure}

A more credible estimate 
of  the relative sensitivity of colliders and
$\tmg$ to the parameter $\kappa_{\tau \mu} \tan \b_\tau$,
could be obtained from a double ratio, as discussed
for $(g-2)$ and $\tmg$ at the end of 
section \ref{sec:tmg}.
We plot in figure \ref{ratiosigtmg} the
double ratio of  
$R^\sigma_h R^{BR}_h/$[the current bound on $h_{SM} \to \tau \bar{\tau}$],
divided by the predicted $BR(\tmg)$  normalised to its
current experimental bound.  Both rates are $\propto
( \kappa_{\tau \mu} \tan \beta_\tau)^2$, so this cancels
out of the ratio. This plot is made
for the production, and decay
to $\tau^\pm \mu^\mp$,  of a CP even Higgs $h$
of mass 115 GeV,  with   $m_H = 130$ GeV,  $m_{H+} = 300$ GeV
and $|m_A - m_{H+}| = 10 $ GeV. 
If $m_A \simeq m_{H+} \gsim 250$ GeV
for B physics and the T parameter, then 
 the current bound from $\tmg$ makes it difficult to
detect  $h \to \tau \bar{\mu}$ at the TeVatron
\footnote{To rescale this plot for the LHC is straightforward:
multiply by the TeVatron limit of $\simeq 30 \times$ SM 
expectation for $h_{SM} \to \tau^+ \tau^-$, and divide
by the LHC limit on $\sigma (gg \to \phi \to  \tau^\pm \mu^\mp)/
\sigma (gg \to h_{SM} \to  \tau^+ \tau^-)$.}.
The sensitivity to $H \to \tau \bar{\mu}$ is
worse, in the interesting 
$\sin (\b - a)  \sim \cos (\b - a) \sim \tan \beta_\tau$ region,
due to sums and differences among these angles all
of the same orders. 
However, a hadron collider able to detect   
$h_{SM} \to \tau \bar{\tau}$ (an improvement
of $\sim 30 $ with respect to the
current Tevatron limit), would be more
sensitive than $\tmg$ in
the  $\sin 2(\b - a) \gsim 1/2 $ and 
small $\tan \beta_\tau$ region. 
Also,  cancellations are possible
in the $\tmg$ amplitude, which would increase the
sensitivity of colliders relative to
rare decays. 

There are dubious approximations in  obtaining
this plot.  We neglect the contribution
of the $\phi \to \tau \bar{\mu}$ decay 
in computing the  total $\phi$ decay rate (this
approximation was  also used to
obtain eqn (\ref{Tevguess})). We make this
overestimate of the branching ratio,
so as  to  obtain a formula which is $\propto  |g_{\phi \tau \mu}|^2$,
so easy to compare to $BR(\tmg)$. It is reasonable
where $\tmg$  imposes $\kappa_{\tau \mu} \tan \beta_\tau \lsim 3$
(because then $\Gamma(\phi \to \tau 
\bar{\mu}) \lsim  \Gamma(h \to \tau 
\bar{\tau})$, and the contribution to the
total decay rate is not to important); however,
in the interesting 
small $\tan \beta_\tau$ and large $\sin 2 (
\b - \a)$ region, it could overestimate
$\sigma (p \bar{p} \to h \to \tau^\pm \mu^\mp$
by a factor $\sim 2$ (see figure \ref{ratio;c2v2}). 
A second  doubtful approximation is
 that the higher order QCD
corrections cancel  in eqn (\ref{ratiosigma}).
This may be  acceptable when 
$\phi$ and $h_{SM}$ are emitted
from a top loop, but in the large $\tan \b_\tau$ limit,
the $b$ loop can also be important for $\phi$
production, and its NLO corrections are
different. 
Finally, the experimental bound we use  from
the search for $h_{SM} \to  \tau \bar{\tau}$\cite{D0}, allows
for several  Higgs production mechanisms:
associated production with a
$W$ or $Z$, vector boson fusion, or gluon fusion,
and assumes a final state of $\tau \bar{\tau} +$ 2 jets.  
Whereas in (\ref{ratiosigma}), we assume
production by gluon fusion. 
 
We can also compare our prediction to bounds
\cite{D0MSSM} on a MSSM Higgs produced via gluon fusion,
 and decaying to  $\tau^+ \tau^-$. This analysis assumed
  one $\tau$ decaying hadronically and the other one 
to a muon and neutrinos. The cross section limits on 
$\sigma \times BR(\phi \to \tau \tau)$ displayed in figure 5 of \cite{D0MSSM}
 indicate  the current TeVatron limit on 
$gg \to \phi \to \tau \mu$. In  $\phi \to \tau^\pm \mu^\mp$, 
the muon is directly produced in the decay of the Higgs, rather 
than from  a $\tau$, so 
the signal  would have a global detection efficiency roughly
5 times higher than the signal of \cite{D0MSSM},
 due to the absence of the factor 0.17
coming from the decay rate of $\tau$ in $\mu$. 
So the y-axis in figure 5 of \cite{D0MSSM}
could roughly be labelled as 
$\sigma \times BR(\phi \to \tau \mu)/0.17$.
For a Higgs of 115~GeV,  this gives a limit around 30 times the SM 
expectation, which is similar
to the limit quoted in \cite{D0} with twice the luminosity.

For experimental bounds, we have extrapolated the potential reach of 
$h \to \tau \mu$ searches from the published searches for $h \to \tau \tau$.
In \cite{Assamagan:2002kf}, it has been shown that by using appropriate cuts, 
it was possible to distinguish between $h \to \tau \mu$ and $h \to \tau \tau$
signals (see figures 13 and 14 in \cite{Assamagan:2002kf}).
Various cuts can help extract a $h \to \tau \mu$ signal from the standard model
backgrounds that are still selected by a $h \to \tau \tau$ analysis. Increasing
the muon Pt threshold is a first option. In \cite{Assamagan:2002kf}, this 
threshold is at 20 GeV, already above the thresholds used in $h \to \tau \tau$
searches (10 GeV for CdF and 15 GeV fo D0 \cite{TEVNPH}). With Higgs mass above 
100 GeV, increasing the threshold up to 30 GeV can safely be done. 
Figure 1 of \cite{D0MSSM} shows that threshold increase would greatly reduce 
the amount of standard model backgrounds remaining in a $h \to \tau \tau$ 
analysis. The dominant remaining backgrounds would be Z$\to \tau \tau$, 
Z$\to \mu \mu$\footnote{not considered in \cite{Assamagan:2002kf}} and W+jets.

Another potentially efficient variable to isolate $h \to \tau \mu$ signal is
the effective transverse mass of the tau mu system as defined in equation (30) 
of \cite{Assamagan:2002kf}. For W and Z background, this variable would peak 
at the W/Z mass while for higgs signal, it would peak at the higgs mass
(see figure 11 of \cite{Assamagan:2002kf}). Hence, a cut on this variable could 
further reduce the background especially for higher higgs masses.

Compared to the $h \to \tau \tau$ analysis, a dedicated $h \to \tau \mu$ analysis 
could then have fewer background events for a similar selection efficiency 
leading to sensitivity and limits which would be better than the one crudely 
extrapolated from current experimental limits on $h \to \tau \tau$ searches \cite{D0,D0MSSM}.

\section{Summary}

At hadron colliders, an interesting signature
of New Physics would be the  production of a  neutral Higgs
 $\phi$, followed by its decay to $\tau^\pm \mu^\mp$.
For $m_\phi \ll 2 m_W$, the branching ratio can be large,
and the  process could arise in a wide variety of models.
Such New Physics  is  constrained by precision
observables, and the upper bounds on rare decays such as 
$\tau \to \mu \gamma$.  In this paper, we are interested
in the implications for  collider searches of
these loop contributions.

We  parametrise
the New Physics at mass scales
$\lsim 400$ GeV,  as a (CP- conserving) 
2 Higgs Doublet Model (2HDM) of Type III, meaning
that we allow our neutral  Higgses to have the  tree-level lepton
flavour changing couplings of eqn (\ref{FVcplings}):
$$
\sim \kappa_{\tau \mu}   \tan \beta_\tau \sqrt{\frac{m_\mu m_\tau}{v^2}}
\left(
 \cbma  \hl
- \sbma  \hh
+i
\ha
\right) \overline{ \tau}  \mu
$$
Our choices of  notation, basis in Higgs space,
$\tan \beta_\tau$,
and parametrisation are discussed in section
\ref{notn}. We will quote and plot bounds on
$ \kappa_{\tau \mu}   \tan \beta_\tau$,  to avoid
artificially strong bounds on $\kappa_{\tau \mu} $
at large $\tan \beta_\tau$.
 We study the phenomenological constraints
on this model:  we admit arbitrary New Physics at
scales $\sim $ TeV, and retain the  contraints
on the 2HDM arising from  the $T$ parameter, $b \to s \gamma$,
$(g-2)_\mu$ and $\tmg$.

The $T$ parameter and $b \to s \gamma$ restrict the
flavour-independent parameters (masses) of the model.
The charged Higgs of the 2HDM gives a significant
same-sign contribution to the  Standard Model (SM) amplitude
for $b \to s \gamma$ (see sect \ref{flav}). For $m_{H+} \gsim 250-300$ GeV,
this contribution is within current experimental error.
In this paper, we assume such a heavy $H^+$;
this has little effect on the neutral Higgses we
are interested in.
The $T$ parameter, discussed in section \ref{secSTU},
is the sum of many terms of different sign. One
way to ensure that it is
small enough  is to take  $m_A \sim m_{H+}$, 
as we  assume  for most of the plots. However,
other cancellations are possible within
the 2HDM (such as  $m_H \sim m_{H+}$ with 
$\sin(\b - \a) \to 1$),  or additional new light particles
(such as arise in supersymmetric models) could
contribute to $T$.

The  decay 
 $\tmg$  and
the anomalous magnetic moment of the muon $(g-2)_\mu$
  constrain the flavour-changing
coupling $\kappa_{\tau \mu} \tan \beta_\tau$,
and are discussed in sections \ref{ad} and
\ref{sec:tmg}.
The neutral Higgses $h,H$ and $A$   contribute to $(g-2)_\mu$
at one loop, with a flavour-changing coupling
at both vertices, and chirality-flip via
a $\tau$ mass  insertion. One could hope
to fit the $(g-2)_\mu$ discrepancy, with a 2HDM
containing the  $\kappa_{\tau \mu}$ coupling,
but this would require cancellations  in the  
generically more restrictive $\tmg$ bound (see
figure \ref{ratio;c1} and  eqns 
(\ref{daapp}), (\ref{bd1loop}) and (\ref{tmg;c1})).

There are relevant contributions to $\tmg$ at
one and two-loop; some diagrams are  shown in figures \ref{bz} and \ref{bzg}.   
The one loop amplitude scales  \footnote{This scaling
is with our choice $\tan \b_\tau$ factors in the
lepton flavour violating coupling. Other
authors \cite{Paradisi:2005tk,Kanemura:2005hr}   take
this coupling $\propto 1/\cos^2 \b$, and find
a branching ratio $\propto \tan^6 \beta$.}
as $ \frac{m_\tau ^2}{v^2}  \tan^2 \b_\tau(\kappa_{\tau \mu} \tan \beta_\tau)^2$,
and does not vanish for $\sin 2(\b - \a) \to 0$.
The  amplitude due to $A$ exchange  is of opposite sign
from the $h$ and $H$ amplitudes, so the one-loop
contribution  to $\tmg$ can be suppressed by
tuning similar masses for $h,H$ and $A$. 
Notice however, that the  tuning required 
becomes finer with increasing $\tan \b_\tau$, so
at  large $\tan \b_\tau$, $\tau$ decays are a 
more sensitive probe
\cite{Kanemura:2005hr} of $\kappa_{\tau \mu} \tan \beta_\tau $
than hadron colliders. See \cite{Kanemura:2009xs} 
for a discussion of the  promising prospects at an $e- \gamma$
collider.

The two-loop ``Barr-Zee'' contributions
to $\tmg$   can be of the 
same order as the one loop. 
The effective dipole interaction of
$\tau, \mu$ and $\gamma$, must contain
at least three Yukawa couplings
at any loop order;  if the Higgs has ${\cal O}(1)$
couplings to the tops, then  two of
the three Yukawa couplings in the two-loop diagram
can be top Yukawas. At large $\tan \b_\tau$, the
$\phi$-induced effective interaction $\bar{t} t\bar{\tau} \mu$
  vanishes with $\sin 2(\b - \a)$. 
In this $ \sin 2 (\b - \a) \to 0$ limit, one  of $h$ and $H$
becomes ``Standard-Model-like'',
with large couplings to the $t$ and $W$,
while  $A$  and the other  CP-even scalar
have unsuppressed flavour-changing interactions.
However,  as illustrated in figure \ref{bz}, the
``Barr-Zee'' diagrams are closely related
to Higgs production by gluon fusion. 
So if the contribution to $\tmg$ is suppressed
in this way, the collider cross-section
is suppressed as well.
The two loop top diagrams
tend to dominate the $\tmg$ amplitude
at small $\tan \b_\tau$ and large 
 $\sin 2(\b - \a)$. Their contribution
can be suppressed by cancellations between
diagrams --- for instance  the $h$ amplitude   cancels against
the   $H$ amplitude 
when $1/\tan \b_\tau \to 0$ and $m_h \to m_H$ --- but these
cancellations do not arise for  the same parameter choices as
suppress the  one loop amplitude. 

So in practise, the  model is described by $m_{H+}$
(taken $\simeq 300$ GeV to satisfy $b \to s \gamma$), 
$m_h (\simeq 114$ GeV), $m_A$ and $m_H$ 
(one of which should be ``light''  to avoid the decoupling limit,
and the  the $T$ parameters
constraint can be satisfied if  the other is heavy),
$\tan \beta_\tau$ ($\sim$ few,
to maximise Higgs production by gluon fusion at the
Tevatron and the LHC), and $\sin (\b - \a)$
(which is unconstrained if $m_A \simeq m_{H+}$,
and is constrained to be $\sim 1$ by
the $T$ parameter if $m_H \simeq m_{H+}$).
Despite the freedom to vary $\sin(\b - \a)$,
we find it difficult to generate a detectable cross-section
at the Tevatron with parameters that satisfy the
precision constraints.

In the absence of cancellations, $\tmg$ is a more
sensitive probe of the  $ \kappa_{\tau \mu}$ coupling,
than $\phi \to \tau^\pm \mu^\mp$ at the TeVatron
with $\sim 5$/fb data. 
This can be seen from figure \ref{ratiosigtmg}.
However, it is  possible to evade  constraints 
 that arise from loops   by tuning parameters, or
by adding (more)  New Physics. We have not
studied this, for reasons of principle and practise.
Our approach is phenomenological, so tuning 
masses and  mixing angles 
to cancel one and two loop contributions
appears ad hoc. In a model,  cancellations can
be justified by symmetries. More practically, 
we do not use complete gauge invariant two
loop formulae, and do not have an estimate for
 higher order effects, so cancellations we find
might not persist in a better analysis.

Direct collider searches can give clearer bounds,
being less dependent than loop
processes  on the vagaries of cancellations. 
Multiplying figure \ref{ratiosigtmg} by the current
Tevatron bound on $\sigma (p \bar{p} 
\to h_{SM} \to \tau^+ \tau^-)$ which is
$\sim 30\sigma_{SM}$,  shows that 
$\sigma (gg \to h \to \tau^\pm \mu^\mp)$
can be of order $\sigma (gg \to h_{SM} \to \tau^+ \tau^-)$,
and consistent with $\tmg$ (for small $\tan \b_\tau$).   
So observing  $\phi \to \tau^\pm \mu^\mp$ in the near future
 would require   ``tuning'' 
by the ratio plotted in  figure \ref{ratiosigtmg}:
a factor $\sim .1$ for small $\tan \beta_\tau$.

\end{document}
\bye